\documentclass[aps,pra,showpacs,twoside,twocolumn,10pt]{revtex4-1}
\usepackage[colorlinks=true, citecolor=red, urlcolor=blue ]{hyperref}
\usepackage{epsfig,newlfont,amssymb,amsfonts,amsmath,bm,subfigure,palatino,mathtools,amsthm,braket,times,soul,enumitem,color}
\usepackage[normalem]{ulem}
\newcommand{\stkout}[1]{\ifmmode\text{\sout{\ensuremath{#1}}}\else\sout{#1}\fi}

\usepackage[utf8]{inputenc}
\usepackage{array}
\usepackage{graphics}
\newcommand{\ketbra}[2]{|#1\rangle \langle #2|}
\begin{document}

%\title{Disorder-induced enhancement of precision in quantum metrology}

%\title{Disorder in quantum device accessory can enhance metrological precision}
%\title{Tunable disorder in accessories can enhance precision of atomic clocks}
\title{Enhancing precision of atomic clocks by tuning disorder in accessories}

\author{Aparajita Bhattacharyya, Ahana Ghoshal, Ujjwal Sen}
\affiliation{%Harish-Chandra Research Institute, HBNI, Chhatnag Road, Jhunsi, Allahabad  211 019, India
Harish-Chandra Research Institute, A CI of Homi Bhabha National Institute, Chhatnag Road, Jhunsi, Prayagraj 211 019, India}

\begin{abstract}
We find that a quantum device having an accessory involving precision measurement can have an enhancement of its metrological precision in estimating an unknown parameter of the quantum system  by insertion of glassy disorder, accidental or engineered. We clearly mention how an unbiased estimator can also be identified in a disordered situation, and how the precision thereof can be bounded by the quantum Cr{\'a}mer-Rao inequality. We compare the Fisher information-based lower bound of the minimum standard deviation of an unbiased estimator, in presence of glassy disorder in the system, with the same of an ideal, viz. disorder-free, situation.  The phenomenon can boost the efficiency of certain measuring devices, such as atomic clocks. The precision of these clocks, when measuring time, hinges on the precise determination of the frequency of a two-level atom. In cases where impurities are present in the atom, and can be modeled as a disorder parameter, it is possible for the measurement of frequency to be more accurate than in an ideal, disorder-free scenario. Moreover, disorder insertion can reduce the requirement of entanglement content of the initial probes, which are copies of two-qubit states, along with providing a disorder-induced enhancement.
\end{abstract}

\maketitle
\section{Introduction}
\label{Sec:1}
%\noindent \emph{\textbf{Introduction.}}- 
In the last few decades, quantum metrology~\cite{Braunstein1,Lloyd0,Treutlein} has become an important area of research and its application in various arenas of physics has been delved into~\cite{q_imaging,Genoni3,Liang,therm1,therm2,gravity,mag1,mag2,Holland,opt7,Dowling,Sanders,Ou,Heinzen,Cirac&Plenio,Maccone,Giovannetti_Maccone,Pirandola,Plenio1,Horodecki,Toth,Lewenstein1,Lukin,Nagata,Roy,Lewenstein,Matsuzaki,Hu,opt1,opt2,opt3,opt4,opt5,opt6,Giovannetti_Maccone2,Cramer,Helstrom,Holevo,Braunstein,Milburn,Hayashi1,Hayashi2,Paris_book,Paris4}. 
For optimal choice of encoded probes and decoding schemes, the error in measurement can be minimized, and that minimum error is achievable by repeating the experiment a large number of times~\cite{Braunstein,review1}. 

 Imperfections, modelled as disorder, are ubiquitous in all practical implementations of a physical process~\cite{Cirac&Plenio,Saunders,Shaji,Genoni,Wang,Binder,Agarwal,Huelga2,Paris3,Kurizki,Huelga,Paris,Paris2,Aharony,Misguich,Villain,Minchau,Henley,Moreo,Feldman,Volovik,Abanin,Adamska,Santos,Benneti,Lakshminarayan,Karthik,Brown,Dukesz,Hide,Fujii,Prabhu,Munoz,Sacha,Sacha2,Wehr,Osborne, Foster,Anderson,Brout,Thomas,Anderson1,Lee,Young,Albert,Goltsev,Ahufinger,Chakrabarti,Fallani,Inguscio,Alloul,Wehr_Sacha,Sanchez,Modugno,Aurbach,Shapiro,Hide_new,Alvarez,Havlin,Pollmann,Yao,Eisert,Fanchini,Prabhu2,Prabhu3,Bera1,Bera2,Aharony1,Bera3,Ghosh1,Ghoshal_Das,Sarkar_Sen,Tanoy,Matsuzaki2,An,Huelga3,Berrada,Shi,Tong_Luo,Matsuzaki_new,Xu,Lin}. For example, let us consider the measurement of time or frequency using Ramsey interferometry in quantum devices, like atomic clocks~\cite{clock1,clock2,clock3,clock4}. These clocks are known for their high precision in measuring time. However, even the most well-designed quantum systems have minor flaws that influence their performance. These imperfections and flaws can be characterized as disordered parameters inherent to the system. Specifically, one notable source of disorder in atomic clocks arises from the regulated probing of atoms. This controlled interaction with atoms, while critical for precise measurements, can change the energy difference between the two energy levels of a two-level system. This variation, in turn, can affect the clock's accuracy and stability. %Particularly, the energy difference between the energy levels of a two-level system can change due to the regulated probing of atoms, which can be physically realisable in cold atom systems.
 %Therefore the response to  imperfections, modelled as disorder, in the system are of prime significance. 
 %Imperfections, modelled as disorder, are ubiquitous in almost all physical processes, and have typically detrimental effect in any quantum task. 
 Therefore, response to disorder in measurement precision for such measuring devices is of prime significance in parameter estimation protocols,  and yet is hitherto unexplored.
%
%
%\textcolor{magenta}{There may be imperfections in measurement while designing instruments like atomic clock, which eventually affects its sensitivity. This can be modelled as}
We consider disorder present 
%a special type of disorder which is present in the Hamiltonian, 
in the scale of the parameter to be estimated.

We construct disorder models for measuring devices, especially for atomic clocks, by introducing disordered parameters - more precisely, effective glassy disordered parameters - into the ideal system Hamiltonian, 
where the disorder distributions are chosen from a clutch of paradigmatic
%where the parameters are chosen from different 
continuous and discrete distributions. In such a disordered system, we find scenarios where the presence of disorder provides an improvement in the efficiency of estimation of an unknown system parameter, over the ideal situation. These situations of attaining better efficiency of measurement in presence of disorder over the ideal scenario can be termed as ``disorder-induced enhancements of metrological precision" or instances of ``order from disorder". 
%%%%%%
Therefore, a quantum device containing an accessory that involves estimating a parameter, such as an atomic clock measuring time or frequency, can attain a better precision when a system parameter is inflicted - accidentally or 
%engineered 
intentionally
- by disorder.
%%%%%%
%The disordered system parameters considered are assumed to be ``glassy".
%%%%%%%%
%Moreover, we show that the presence of glassy disorder in a system can be advantageous, while measuring a system parameter over the ideal case, in the aspect of quantum metrology also.
Moreover, we find scenarios where for an ideal system, the best metrological precision is obtained for maximally entangled initial probes, whereas insertion of disorder into the system provides a \emph{double advantage}: an enhancement of metrological precision over the ideal one, and  lower entanglement content in the probe state.
%We estimate the error in the system frequency using the expression of quantum Fisher information, in presence of disorder, and observe that for specific probability distributions from which the disorder parameter is chosen, we get benefit over the disorder-less scenario. We take into account single and two qubit cases. Additionally in the two qubit picture,  we consider interacting systems in presence of parallel field, where we estimate the coupling constant, and obtain an ``order from disorder" phenomenon also in this situation.
%Another intriguing feature in presence of interactions occur as the quantum correlations between the two qubits play a significant role. It is observed that, while the ``order from disorder" takes place, the average entanglement between the two qubits of the optimal input state swtiches from $1$ to a much lower value when disorder is turned on. This is noted for a specific choice of coupling constants. So we attain a metrological advantage in presence of disorder, with a less requirement of two qubit correlations considered as a resource.

For dealing with metrological precision in disordered scenarios, we analyze how an unbiased estimator can  be identified in a disordered situation. We go on to further show that  the precision thereof can be bounded by a quantum Cr{\'a}mer-Rao inequality.

%\textcolor{blue}{
%For $N$ independent parties and non-unitary encodings, the maximum scaling that can be achieved in the minimum error in the estimation of a channel parameter, is an improvement of $\mathcal{O}(1/\sqrt{N})$ by a constant factor, both in the asymptotic limit of $N$~\cite{Rafao_dis1} and in the finite-$N$ limit~\cite{Rafao_dis2}.
%It is known that the maximum scaling in presence of noise can be better than $\mathcal{O}(1/\sqrt{N})$, but not whether it can achieve the Heisenberg limit is still unexplored.
%See also~\cite{Cirac&Plenio,Maccone,noise0,noise1, noise2, noise3,decoher1,decoher2,decoher3,decoher4,decoher5,decoher6,decoher7,Campo,noise4} in this regard.
%In our work however, we restrict only to unitary encodings in presence of glassy disorder, in contrast to the previous works where  decoherence has been considered. Limiting to only product inputs, we find that the disordered situation is advantageous over the ordered one, with an enhancement 
%of the $1/\sqrt{N}$ precision 
%by a factor depending on the type of distribution from which we choose the disorder parameter.
%So while the results in~\cite{Rafao_dis1} pertain to arbitrary inputs - which may include entangled ones- and non-unitary encodings, our work considers product inputs and unitary encoding, but in presence of glassy disorder. We consider arbitrary inputs only when the encoding Hamiltonian contains interaction terms.}
It is important to mention here that our work falls within the general category of imperfections in the encoding process. Until now, this has been considered using noisy quantum operations~\cite{Cirac&Plenio,Maccone,Rafao_loss,Rafao_dis1,Rafao_dis2,Rafao2,Rafao3,noise0,noise1,noise2,noise3,decoher1,decoher2,decoher3,decoher4,decoher5,decoher6,decoher7,Campo,Campo2,R_emni,emni1,emni2,new0,Matsuzaki,new1,noise4,
%}. See also~\cite{
ref1,ref2,ref4,ref5,ref6,ref7}. 
%in this regard. 
We however analyze its implementation by considering disorder present in the encoding unitary. 
Moreover, while the results such as in~\cite{Rafao_dis1} pertain to arbitrary multi-party inputs - which may include entangled ones - and non-unitary encodings, our work uses copies of single-qubit states and unitary encoding, but in presence of glassy disorder. We consider arbitrary multi-party inputs only when the encoding Hamiltonian contains interaction terms.

The rest of the paper is presented as follows. In Sec.~\ref{Sec:2}, we briefly discuss certain aspects of quantum metrology.
%and the measurement scheme that we have taken in this paper. 
The concept and origin of glassy disorder and  disorder averaging are discussed in Sec.~\ref{Sec:3}. The  disorder distributions, that provide disorder-induced enhancement of metrological precision, are also presented in this section. 
In Sec.~\ref{extra}, we identify an unbiased estimator for the disordered situation considered, and show that the precision thereof can be bounded by the Cr{\'a}mer-Rao inequality. 
%refers to the probability distributions required for our calculation of the quench averaged error in the estimated parameter.
Sections~\ref{Sec:5} and~\ref{Sec:6} constitute evidence of order from disorder scenarios for copies of uncorrelated and two-qubit maximally entangled initial probes, for different glassy disorder models. The reduction of the requirement of entanglement content of the initial probes in presence of glassy disorder, thus providing the ``double advantage'', is shown in Sec.~\ref{Sec:7}.
%Sec.~\ref{Sec:5}, we further investigate the two qubit scenario in presence of interactions between the two qubits and application of parallel field. 
A conclusion is presented in Sec.~\ref{Sec:8}.

%\vspace{0.2cm}

%\noindent
%\textit{\bf{Estimation of an unknown parameter.}}- 

\section{Estimation of an unknown parameter}
\label{Sec:2}
The enhancement of measurement precision in the frequency estimation protocol using the concepts of quantum metrology has been studied some time back in~\cite{Cirac&Plenio,Maccone}. The problem initiated with a single-qubit system described by the Hamiltonian $\mathcal{H}_1$, given by
\begin{equation}
\label{H0}
  \mathcal{H}_1=-\hbar\omega \ket{1}\bra{1}, 
\end{equation}
where $\omega$ is the frequency to be estimated. For an $n$-qubit system, the dynamics is governed by the $n$-qubit Hamiltonian,
\begin{eqnarray}
    \mathcal{H}_{n} &=&\mathcal{H}_1\otimes \mathcal{I}_2\otimes\mathcal{I}_3\cdots \otimes \mathcal{I}_{n}+\mathcal{I}_1\otimes \mathcal{H}_1\otimes\mathcal{I}_3\cdots \otimes \mathcal{I}_{n}\nonumber\\
    &&\phantom{aji dhaner}+\cdots +\mathcal{I}_1\otimes\mathcal{I}_2\cdots \otimes\mathcal{I}_{n-1} \otimes \mathcal{H}_1.
\end{eqnarray}
For the evolution of $n$-qubit initial probes by the unitary $U_n=e^{-\frac{i}{\hbar} \mathcal{H}_n t}$, the minimum standard deviation in the estimation of $\omega$, is less for two-qubit maximally entangled probes than for copies of uncorrelated ones. This enhancement of metrological precision by introducing quantumness in the initial states is a key idea of quantum metrology.

The minimum deviations in frequency estimation
%\textcolor{blue}{, given in Eq.~(\ref{nl_prod}),}
are not due to imprecise
%the actual error in 
 measurements of 
 %\textcolor{blue}{$\omega$} 
 the parameter in an experiment.
It is a fundamental lower bound on the standard deviations of estimators %$\hat{\theta}(x)$, 
of the parameter 
%$\theta$ 
to be estimated, where the outcomes of measurements - on which the estimators depend - 
%$x$, 
are distributed according  to the  probabilities  dictated by the Born rule.
%of obtaining a particular outcome.
%density function
%, $f(x|\theta)$. 
This is the quantum Cram\'{e}r-Rao bound~\cite{Helstrom,Holevo,Braunstein,Milburn,Hayashi1,Hayashi2,Paris_book,Paris4}, 
%which, for unbiased estimators,
%
%in which there is a tacit assumption that the estimators are unbiased, 
%i.e., when $\langle \hat{\theta}(x)\rangle = \theta$,} 
%The lower bound corresponding to the optimal measurement is referred to as the quantum Cram\'{e}r-Rao bound, given by}
%The quantum Cram\'{e}r-Rao bound 
which is
 %\textcolor{blue}{It is a fundamental bound on the standard deviation of estimators, $\hat{\theta}(x)$, of the parameter $\theta$, where the outcomes of measurements, $x$, belong to the probability density function, $f(x|\theta)$, given by the Cram\'{e}r-Rao bound~\cite{Helstrom,Holevo,Braunstein,Milburn,Hayashi1,Hayashi2,Paris_book,Paris4}. This lower bound is}
% \textcolor{red}{\sout{It is the lower bound of the deviation of the estimated parameter, which directly comes from the quantum Cram\'er-Rao bound and is}} 
 attainable for a large number of repetitions of an experiment corresponding to an optimal strategy.
%in the asymptotic limit ($\nu \rightarrow \infty$). 
Below we discuss the prescription of obtaining 
this
%the quantum Cram\'er-Rao 
bound.
Suppose an unknown parameter, $\theta$, is encoded in a state, $\rho(\theta)$, of a physical system. To estimate $\theta$, a measurement of elements $\{\Pi_x\}$ is done on the system, and let the probability distribution for a measurement outcome $x$ be $f(x|\theta)$. Now, let us take the outcome of a single measurement to be $x_1$. Based on this outcome, the predicted value of $\theta$ can be represented by the estimator function $\hat{\theta}(x_1)$~\cite{Maccone,review1}. For a large number of measurements, our prediction is assumed to be correct on an average for a fixed $\theta$. Therefore,
\begin{equation}
    \langle \hat{\theta}(x) \rangle_\theta = \int dx f(x|\theta) \hat{\theta}(x) = \theta. 
\end{equation}
Such a $\hat{\theta}(x)$ is called an unbiased estimator. So, for a given $\theta$, we get a distribution of the estimate $\hat{\theta}(x)$ with probability $f(x|\theta)$. The variance of the estimator function, while estimating $\theta$, is lower bounded by the Cram\'{e}r-Rao inequality given by
\begin{equation}
\label{ClCramerRao1}
    \Delta^2 \theta \geq \frac{1}{F(\theta)},
\end{equation}
where $F(\theta)$ is the Fisher Information (FI)~\cite{Braunstein,review1}. Here we use the traditional notation \(\Delta^2 \theta\) for 
\(\Delta^2 \hat{\theta}(x)|_\theta\) as the variance of the distribution of the estimator function \(\hat{\theta}(x)\). The Fisher information for a single measurement is defined in terms of the probability distribution $f(x|\theta)$, through the relation
\begin{eqnarray}
F(\theta) &=& \int dx \big[\frac{\partial}{\partial \theta}\log f(x|\theta)\big]^2.
\label{FI}
\end{eqnarray}
From the additivity of FI, for $\nu$ measurements, or $\nu$ copies of the input probe, we have the FI, $F^{\nu}(\theta)=\nu F(\theta)$, where $F(\theta)$ represents the Fisher information corresponding to each probe.
%If the number of measurements for a fixed measurement strategy be $\nu$, the Fisher Information, $F^{\nu}(\theta)$, can be evaluated using
%\begin{eqnarray}
%\label{F_nu}
%F^{\nu}(\theta) &=& \int dx_1 dx_2...dx_\nu \big[\frac{\partial}{\partial \theta}\log\big( f(x_1|\theta)f(x_2|\theta)...f(x_\nu|\theta)\big)\big]^2, \nonumber \\ 
%&=& \nu F(\theta) \nonumber
%\end{eqnarray}
%where integration is performed across all the measurement outcomes.
%$F^{\nu}(\theta)=\nu F(\theta)$.
Therefore for $\nu$ measurements, ~(\ref{ClCramerRao1}) has the general form 
%of the Classical Cram\'{e}r Rao inequality and this bound is tight in the asymptotic limit$(\nu \rightarrow \infty)$ by the maximum likelihood estimator.
\begin{equation}
\label{ClCramerRao}
    \Delta^2 \theta \geq \frac{1}{\nu F(\theta)},
\end{equation}
which attains the equality for large $\nu$.
%in the asymptotic limit $(\nu \rightarrow \infty)$. 
The choice of measurement strategy has been fixed so far. To make the FI independent of such a choice, one has to maximize the Fisher information with respect to all possible measurements. As a consequence of the maximization, the Cram\'{e}r-Rao bound becomes
%Since the classical Cram\'{e}r Rao bound (Eq.~(\ref{ClCramerRao1})) is  dependent on the measurement approach used, the Fisher information must be maximised with regard to all feasible measurements in order to make it measurement independent and provide the lowest value for the bound.
%Hence the the quantum Cram\'er Rao bound takes the form,
\begin{equation}
\label{QCRbound}
\Delta \theta \geq \frac{1}{\sqrt{\nu \big(\max_{\Pi_x} F(\theta)\big)}} = \frac{1}{\sqrt{ \nu F_Q(\theta)}}.
\end{equation}
$F_Q$ is the quantum Fisher information corresponding to a single probe, obtained by maximizing the Fisher information $F(\theta)$ with respect to all possible  $\{ \Pi_x\}$, where $f(x|\theta)=\text{Tr}[\Pi_x \rho(\theta)]$. Here $\{ \Pi_x\}$ is a positive operator valued measurement and $\rho(\theta)$ is the quantum state where $\theta$ is enoded.
%In case of quantum encoding and decoding let, $f(x|\theta)=\text{Tr}[\Pi_x \rho(\theta)]$, where $\{ \Pi_x\}$ is a positive operator valued measurement and $\rho(\theta)$ is a quantum state. 
The quantum Fisher information can also be directly calculated for a given $\rho(\theta)$ as 
\begin{equation}
\label{QFI}
F_Q(\theta)=\text{Tr}\Big[\rho(\theta) {L_s\big[ \rho(\theta) \big]}^2 \Big].
\end{equation}
%and the corresponding Cram\'er-Rao bound is referred to as quantum Cram\'er-Rao bound. 
 $L_s\big[ \rho(\theta) \big]$ is known as the symmetric logarithmic derivative (SLD)~\cite{Braunstein,Maccone} of $\rho(\theta)$. 
%and is defined via the equation,
%\begin{equation}
%    \frac{\partial \rho(\theta)}{\partial \theta}=\frac{1}{2}\Big[L_s \big[ \rho(\theta) \big] \rho(\theta)+\rho(\theta) L_s \big[ \rho(\theta) \big]\Big].
%\end{equation}
If the eigen-decomposition of $\rho(\theta)$ is $\rho(\theta)=\sum_i \lambda_i(\theta) \ket{e_i(\theta)}\bra{e_i(\theta)}$, the SLD is given by
\begin{equation}
    L_s \big[ \rho(\theta) \big] = \sum_{i,j}\frac{2 \bra{e_i(\theta)} \frac{\partial \rho(\theta)}{\partial \theta} \ket{e_j(\theta)}}{\lambda_i(\theta)+\lambda_j(\theta)} \ket{e_i(\theta)}\bra{e_j(\theta)},
\end{equation}
with the sum being over all pairs of $i$ and $j$ for which $\lambda_i(\theta)+\lambda_j(\theta)\ne 0$.
The maximum of $F(\theta)$ in (\ref{FI}) is obtained by considering the projective measurement in the eigenbasis of the SLD~\cite{Braunstein,Milburn,Rafao1,Rafao2}. For the parameter $\theta$ encoded in a pure state $\ket{\psi(\theta)}$, QFI simplifies to
\begin{equation}
\label{pure_fq}
     F_Q(\theta)=4 \big[\braket{\dot{\psi}(\theta)|\dot{\psi}(\theta)}-|\braket{\dot{\psi}(\theta)|\psi(\theta)}|^2 \big],
\end{equation}
%The classical Cram\'{e}r Rao bound in Eq.~(\ref{ClCramerRao1}) is achievable in the asymptotic limit when $\nu \rightarrow \infty$ and there always exists a measurement strategy where the limit of Eq.~(\ref{QCRbound}) concurs with Eq.~(\ref{ClCramerRao1}). This optimum measurement scheme is provided by the projective measurement in the eigenbasis of SLD~\cite{Braunstein,Rafao1,Rafao2,Rafao3}. As the classical Cram\'er Rao bound (Eq.~(\ref{ClCramerRao1})) is attainable in the asymptotic limit, so is the quantum Cram\'er Rao bound (Eq.~(\ref{QCRbound})) if a suitable measurement strategy is selected.
where $\ket{\dot{\psi}({\theta})}$ refers to the derivative of $\ket{\psi({\theta})}$ with respect to $\theta$~\cite{Rafao2,Rafao3}.
%In our estimation, we have dealt with pure states only, and hence used this expression to calculate the Fisher information.

The minimum uncertainty in measuring a parameter, obtained so far, is optimal with respect to all possible measurements, but still depends on initial choice of inputs $\rho_0^{\otimes \nu}$. So, an optimization over the initial states also has to be done in order to possibly increase the accuracy of the measurement~\cite{Maccone,review1}. Hence, the minimum deviation in the estimation of $\theta$ takes the form
\begin{equation}
\Delta \theta_{opt}  = \min_{\rho_0^{\otimes\nu}} \frac{1}{\sqrt{ \nu F_Q(\theta)}}.
\label{opt}
\end{equation}
Here $F_Q(\theta)$ refers to the quantum Fisher information for a single copy of the input probe, which can be either product or entangled, and $\nu$ is the number of copies of that probe. The number of times the whole estimation is repeated is considered to be one.

%Let us consider $\nu$ copies of the input state where each copy is $N-$party entangled $\ket{\psi^{(N)}_0}$ and unitarily evolves through the equation $\ket{\psi^{(N)}_{\theta}}^{\otimes \nu}=U(\theta) \ket{\psi^{(N)}_0}^{\otimes \nu}$, where the unitary evolution is given by the relation $U(\theta)=e^{-\frac{i}{\hbar}H\theta}$. In such a scenario, the optimum input state~\cite{Maccone} for attaining the bound in the asymptotic limit is of the form 
%\begin{equation}
%\label{best_input}
%    \ket{\psi_{\theta}^{(N)}}=\big( \ket{\lambda_{\max}}^{\otimes N} + \ket{\lambda_{\min}}^{\otimes N} \big)/\sqrt{2},
%\end{equation}
%$\ket{\psi_{\theta}^{(N)}}=\big( \ket{\lambda_{\max}}^{\otimes N} + \ket{\lambda_{\min}}^{\otimes N} \big)/\sqrt{2}$, 
%where $\ket{\lambda_{\max}}$ and $\ket{\lambda_{\min}}$ are the eigenvectors associated with the highest and lowest eigenvalues of the corresponding Hamiltonian $H$.

%If the parameter to be estimated $\theta$ be encoded in a pure state, $\rho(\theta)=\ket{\psi_{\theta}}\bra{\psi_{\theta}}$, the quantum Fisher information in Eq.(~\ref{QFI}) simplifies to

In some situations, the optimal state is easy to determine. For example, let the parameter $\theta$ be encoded in the state through a unitary operator of the form $U(\theta)=e^{-i\tilde{H}\theta}$, with $\tilde{H}$ being the generator of the unitary, independent of $\theta$. In this situation, it can be shown that the best input state for attaining the minimum error in $\theta$ is of the form 
%the quantum Fisher information comes out to be independent of $\theta$ as $F_Q(\theta)=4\Delta^2\tilde{H}$. Hence, the Cram\'er-Rao bound turns out to be
%\begin{equation}
%    \Delta^2\theta \ge \frac{1}{4\nu\Delta^2\tilde{H}}.
%\end{equation}
%This inequality states that in the asymptotic limit, $\Delta^2\theta$ can be minimized by maximizing $\Delta^2\tilde{H}$ .The maximum of $\Delta^2\tilde{H}$ may be found by selecting the initial state as 
\begin{equation}
\label{best_input}
   \ket{\psi({\theta})}_{opt}=\big( \ket{\lambda_{\max}} + \ket{\lambda_{\min}} \big)/\sqrt{2},
\end{equation}
where $\ket{\lambda_{\max}}$ and $\ket{\lambda_{\min}}$ are the respective eigenvectors associated with the highest and lowest eigenvalues of the generator $\tilde{H}$~\cite{Maccone,review1,Bhattacharyya}.
%~\cite{Maccone,review1,Bhattacharyya}. 
Additionally, for the unitary $U(\theta,\phi)=e^{-i(\tilde{H}_1\theta+\tilde{H}_2\phi)}$, where $[\tilde{H}_1,\tilde{H}_2]=0$, while estimating $\theta$ or $\phi$ independently, the best choice of input states remains the same as in Eq.~(\ref{best_input}), with the eigenvectors corresponding to $\tilde{H}_1$ and $\tilde{H}_2$ respectively. For a situation where  $\tilde{H}_1$ and $\tilde{H_2}$ do not commute, the best input state cannot be obtained following this scheme.

%In our prescription, however, we refrain from using the expression of probability \big(Eq.(~\ref{F_nu})\big) to calculate $\Delta \tilde{\omega}$. Our approach is based on the evaluation of the quantum Fisher information using SLD since we are interested in the asymptotic limit i.e. for large $\nu$. Therefore we obtain the expression of the variance of $\tilde{\omega}$ from the qFI which again is computed using the Standard Logarithmic Derivative (SLD).

%We consider the insertion of a disorder parameter, into a single qubit system described by the Hamiltonian $\mathcal{H}_1$. We incorporate the disorder parameter in two ways into the ``ideal" Hamiltonian. The altered forms of the Hamiltonian, with disorder, are taken as
%\begin{eqnarray}
%   &&H_1 = -\hbar\omega(1+\epsilon)\ketbra{1}{1} 
%   \label{eijey}\\
 %  &&\text{and} \quad H_2 = -\hbar(\omega+\epsilon)\ketbra{1}{1}, 
%\end{eqnarray}
%where the disorder is modeled by the parameter $\epsilon$, selected randomly from several paradigmatic distributions, both discrete and continuous. 
%We consider the probability distribution functions, some of which are discrete and some are continuous, and investigate whether they can offer a disorder-induced enhancement in estimation of a parameter over their ideal scenario. 
%Among a few disorder models of different probability distributions, 
%We will find that the metrological estimation provides order from disorder for certain cases.

In this paper, we focus on the minimum error, as given by the right hand side of~(\ref{opt}), in the estimation of a parameter, say $\theta$, by obtaining the QFI directly from Eq.~\eqref{QFI} and choosing the optimal input states following the 
%above mentioned 
prescription around Eq.~($13$) of the same. This actually presents the Fisher information-based lower bound 
%(R.H.S. of Eq.~(\ref{ClCramerRao}))
for the deviation in estimation of an unknown parameter $\theta$, where the measurement basis is taken to be the eigenbasis of SLD, which is the optimal choice of basis. In further discussions, we will only concentrate on this Fisher information-based lower bound, with optimal choice of basis and inputs, and this quantity will be denoted as $\Delta \tilde{\theta}$, for the parameter $\theta$.

\vspace{0.6cm}

\section{Glassy disorder: Origins, models and  averaging}
\label{Sec:3}
For acquiring better precision in metrological estimations, an experiment should be repeated a large number of times to attain the Cram\'er-Rao lower bound. The high number of repetitions of the measurements, in turn, can cause some fluctuations or imperfections in the system parameters which may drive the system some ``distance" away from its ideal nature. Also, as a consequence of the failure to construct an ideal experimental setup, disorder can appear in a system while tuning the parameters of the setup. Irrespective of the origin of the disorder, it may affect the efficiency of a measurement scheme, while estimating an unknown parameter. In this paper, we consider  scenarios, where some types of disorder are present in a system in the form of randomness,
incorporated into the parameters of the system. %while considering only unitary evolution of the optimal pure input state obtained using Eq.(~\ref{best_input}).
%%%%%%%%%%%%%%%%%%%%%%%%%%%%%%%%%%%%%%%%%%%%%%%%%%%%%%%%%%%%%%%%%%%%%%%%%
We take the disorder parameter to be glassy, which means that
%The aim of this paper is to find the effect of glassy disorders present in a quantum system in estimating the error of certain system parameters.
the equilibration time of the disorder in the system is several orders of magnitude larger than the relevant observation time. 
%This indicates that for 
For a specific realization (configuration) of the disorder, the disordered parameters virtually do not alter during the time of investigation~\cite{Chowdhury,Parisi,Dutta,Nishimori,Giardina,Sachdev,Suzuki}. 
%\textcolor{red}{\sout{
They may change after a long time, but that time span is not in our range of interest. Such ``glassy'' systems 
have 
%are usually referred to in the literature as ``glassy"~\cite{Chowdhury,Parisi,Dutta,Nishimori,Giardina,Sachdev,Suzuki,median1,median2}. They have 
also been called ``quenched" disordered systems~\cite{Prabhu2,Bera2,Bera3}. In such scenarios, the disorder averaging must be performed after evaluation of the relevant physical quantity, which,
%has been already computed. 
in our case, 
%we want to compute 
is the Fisher information-based lower bound of the minimum uncertainty in measuring the 
parameter to be estimated,
%frequency, 
in presence of glassy disorder in the system. 
%To achieve this, we first determine the uncertainty in measuring frequency encoded in a physical system, considering an arbitrary but fixed time and configuration of the disorder. The disorder-averaged deviation in frequency is then given by an averaging of the uncertainty in frequency measurement with respect to the disorder distribution. This methodology allows us to derive the disorder-averaged Fisher information-based lower bound.
 A similar averaging is performed for calculating 
 %physically relevant 
 quantities like magnetization and entanglement in glassy systems~\cite{Bera1,Bera2}.
Further details of
%For 
the origin of disorder in physical systems,
the different disorder models and the disorder-averaging method for glassy disorder are provided in the succeeding subsections.
%%%%%%%%%%%%%%%%%%%%%%%%%%%%%%%%%%%%%%%%%%%%%%%%%%%%%%%%%%%%%%%%%%%%%%
%We take the disorder parameter to be glassy, which means that
%The aim of this paper is to find the effect of glassy disorders present in a quantum system in estimating the error of certain system parameters.
%the equilibration time of the disorder in the system is several orders of magnitude larger than the relevant observation time. This indicates that, for a specific realization of the disorder, the disordered parameters virtually do not alter during the time of investigation. They may change after a long time, but that time span is not in our study of interest. Such systems are usually referred to in the literature as ``glassy". They have also been called ``quenched" disordered systems.

%One is the introduction of disorder in the frequency appearing in the Hamiltonian ($H_1$ and $H_2$). This is physically reasonable because the frequency which can applied by a Ramsey pulse may fluctuate due to imperfections in the instrument. 
%The other one is the application of a transverse magnetic field $(H_3)$.
%Eq.(~\ref{H0}) is modified in presence of disorder in the system and the Hamiltonians acquire the form
%\begin{eqnarray}
%later part of our paper, we also take into account two qubit Hamiltonians, where disorder is introduced in both the qubits.

\subsection{How can glassy disorder appear in a physical system?}
Let us begin by examining the origin of such disorders in certain physical measuring devices, specifically in the context of atomic clocks. The precision of time measurement in an atomic clock relies on accurately estimating the frequency of a two-level atom, employing the principle of Ramsey interferometry. An atomic clock functions essentially as an oscillator, where its frequency is tuned to match the transition frequency of a two-level atom. This tuning allows for the precise measurement of the number of oscillations of the atom, effectively defining one second. To delve deeper into this concept let us consider a beam of identical two-level atoms described by the Hamiltonian $\mathcal{H}_1$, emerging from an oven~\cite{clock}. The energy difference between their levels is $\hbar\omega$, with $\omega$ being an unknown parameter. This atom beam traverses between two points, $A$ and $B$, within a waveguide, subject to two microwave pulses originating from a source with a frequency $\omega_f$. During the interaction of the atom with the microwave pulse at point $A$, a coherent superposition of the excited $(\ket{0})$ and ground ($\ket{1}$) states is generated, each with equal weight. The atom then evolves freely up to point $B$ in time $t$, creating a phase difference of $\delta \phi = (\omega - \omega_f) t$ between the atomic superposition and the field.
%This atom beam crosses two microwave pulses, say from point $A$ to $B$ of a waveguide fed by a microwave source whose frequency is $\omega_f$. 
%Due to the interaction of the atom with the microwave pulse at $A$, a coherent and equally weighted superposition of $\ket{0}$ and $\ket{1}$ is generated. Then the atom evolves freely up to $B$ in time $\Delta t$, while a phase difference of $\delta \phi=(\omega-\omega_f)\Delta t$ is created between the atomic superposition and the field. 
At point $B$, a second microwave pulse is introduced, and the detector, placed after the point $B$, measures the probability of the atom being in the ground state of $\mathcal{H}_1$, denoted as $p = (1 + \cos(\delta \phi))/2$.
%The second microwave pulse is fed at point $B$, after which the probability of the atom being in the ground state of $\mathcal{H}_1$, i.e. $p=(1+\cos(\delta \phi))/2$ is being detected at the detector. 
%The oscillator frequency, $\omega_f$, is tuned such that the transition probability, $p$, is maximum, which implies $\omega_f=\omega$. In this manner, one can estimate an unknown frequency gap of a two-level system, which can be further utilized to count the number of atomic oscillations. This sets the standard for defining the unit of one second. Once tuned, the oscillator can be used to measure time in terms of the number of atomic oscillations.
To achieve the maximum transition probability $p$, the oscillator frequency $\omega_f$ is adjusted to obtain the match, $\omega_f = \omega$. This procedure enables us to estimate the unknown frequency gap of the two-level system, which can, in turn, be used to determine the number of atomic oscillations, thus setting the standard for defining one second. 
%After tuning the oscillator, it can be employed to measure time in terms of these atomic oscillations.
However, 
%it is worth noting that 
due to controlled probing of atoms, there may exist tunable disorder within the atom, altering the energy gap between the two levels to $\hbar\omega(1+\epsilon)$ or $\hbar(\omega+\epsilon)$, so that the altered 
%can be described by the 
Hamiltonians are $H_1$ or $H_2$ respectively, where $\epsilon$ is a small 
%\textcolor{blue}{\sout{unknown}}
parameter, 
%\textcolor{red}{
which varies according to some known probability distribution.
%}. 
%\textcolor{blue}{\sout{Although the precise value of $\epsilon$ is unknown, we are aware of the distribution it follows.}}
%\textcolor{red}{
We consider the disorder parameter $\epsilon$, to be fixed for a particular trial of experiment, but selected randomly from a known distribution. It varies unpredictably from one trial to another, but once realized, it is known and fixed for that trial, and is distributed according to a known probability density.
%}
In this paper, we show that, the presence of impurities modelled as $H_1$, can, in certain instances, reduce the error in estimating $\omega$, consequently enhancing the accuracy of frequency estimation. 
%due to the presence of this impurity, the error in the estimation of $\omega$ will be reduced, hence enhancing the accuracy in frequency estimation. 
Consequently, the tuning of the frequency of the oscillator with the beam of atoms that have impurities, will be more precise. If we now connect this finely-tuned oscillator to an ideal, disorder-free beam of atoms, it will accurately count the oscillations, even though impurities were initially present in the atoms during the frequency 
%tuning 
estimating
process.  
%the actual energy gap of the two-level atom can be estimated more precisely in presence of this disorder, thus increasing the precision in measuring the atomic oscillations. After tuning the frequency, if the probing of atoms be withdrawn, then the previously estimated atomic oscillations will give an efficient estimation of time.
%%%%%%%%%%%%%%%%%%%%%%%
Such disordered systems can be effectively modeled in several physical systems~\cite{opt_lat1,ultrac4}. Let us consider the possibility of implementing it in ultracold gas systems~\cite{ultrac4,ultrac1,ultrac2,ultrac3}, which also offers a variety of options, of which we utilize the choice of realizing it in composite fermions formed by fermions and bosonic holes~\cite{opt_lat1,ultrac3}. The effective Hamiltonian in this system contains nearest-neighbor hopping and spin-spin interaction (of strengths \(d_{ij}\) and \(K_{ij}\) respectively), and onsite energy terms (of strength \(\mu_i\)). The \(d_{ij}\) and \(K_{ij}\) are proportional to \(J^2/V\), where \(J\) and \(V\) are respectively the tunneling rate and  onsite inter-particle interaction strength of the parent Bose-Fermi Hubbard model. In the limit of small \(J^2/V\), one is left with an effective Hamiltonian of spinless fermions that is a sum of several decoupled single-particle Hamiltonians of the form in Eq.~\eqref{ham}, with the \(\epsilon\) being independently and identically distributed from one site to the other. Performing interference experiments with each site of that physical system will potentially lead to the desired data required for obtaining a disorder-averaged minimum deviation in frequency estimation, and a potential demonstration of the disorder-enhanced precision in atomic clocks. Note that interferometry with systems of ultracold gas systems is a reality. See e.g.~\cite{matter_wave} for a recent reference. \\

\subsection{Models of disorder} 
\label{sec:3-a}
%some of them discussed below.\\
%, two of which are discrete and three are continuous. In the subsequent sections we will discuss these probability distribution functions which are found to be beneficial over the ideal one.\\
%\\
 %We have considered here a number of truncated distributions for the single qubit and two qubit cases.
 %A conditional distribution generated by limiting the domain of another probability distribution is known as a truncated distribution. Truncated distributions appear in practical situations when the number of events of a probability distribution is limited to values that are above or below a specific threshold or range.
%Precisely the glassy disorders are chosen to be of the following forms:

We enumerate below the models of glassy disorder that we will consider in this paper. In each case, the ``strength" of the disorder is ascertained by the standard deviation of the probability distribution. In the case of the Cauchy-Lorentz disorder, the mean and hence the standard deviation does not exist. The strength in such a case can be defined by using the semi-interquartile range of the distribution.\\

\textbf{\textbullet $\mathbf{\;\;G}$: \textit{Gaussian disorder.}} In this scenario, the disorder parameter, $\epsilon$, is chosen from a Gaussian distribution with a mean of zero and a standard deviation of $\sigma_G$, and the corresponding probability density function is given by
    \begin{equation}
        P(\epsilon) = \frac{1}{\sigma_G\sqrt{2\pi}} e^{-(\epsilon/\sigma_G\sqrt{2})^2}, \;\;\;\; -\infty \le \epsilon \le \infty.
   \end{equation}
\textbf{\textbullet $\mathbf{\;\;CL}$: \textit{Cauchy-Lorentz disorder.}} This disorder is quite different from the others as the mean does not exist for this distribution. The disorder parameter $\epsilon$ is distributed as
\begin{equation}
\label{eq:A1}
     P(\epsilon) = \frac{1}{\gamma \pi }\Big(\frac{\gamma^2}{\epsilon^2+\gamma^2} \Big), \;\;\; \text{where}  \;\;\; -\infty < \epsilon \le \infty .
\end{equation}
%\sout{For uncorrelated initial probes, the disorder-averaged error in frequency estimation diverges for this distribution.}} 
%To gauge the ``strength" of the disorder introduced, we will make reference to the dispersion of the distribution of a disordered system parameter, as quantified by the standard deviation $\sigma$.
As mentioned above, the mean and standard deviation do not exist for this distribution. The strength of the disorder for this distribution is discussed in the succeeding subsection.

\textbf{\textbullet $\;\;\mathbf{D_1}$: \textit{a discrete disorder.}} For this instance, the disorder parameter $\epsilon$ is distributed as
    \begin{eqnarray}
    \label{eq:D1}
    P(\epsilon) &=& \frac{(1-p)}{2}\quad \text{when} \quad \epsilon=\pm\alpha, \nonumber \\
                &=& p  \quad \quad \phantom{tum} \text{when} \quad \epsilon=0, \nonumber \\ %\nonumber %\\
               &=& 0 \; \; \; \; \; \;\;\;\;\;\;\;\;\;  \text{otherwise}, 
    \end{eqnarray}
    with $\alpha>0$ and $0 \le p < 1$. 
    %\textcolor{red}{This distribution is physically meaningful when $p$ is very close to unity.}
%The ``strength" of the disorder is ascertained by 
The standard deviation of the probability distribution is given by
    \begin{equation}
    \sigma_{D_1}=\alpha \sqrt{1-p}.
    \end{equation}\\
    \\
    \textbf{\textbullet $\;\mathbf{D_2}$: \textit{a discrete disorder.}} This is another  discrete distribution which is somewhat different from $D_1$. Here the $\epsilon$ is distributed as
    %For a positive $\alpha$ in the second case, $\epsilon$ is distributed as
    \begin{eqnarray}
    P(\epsilon) &=& \frac{1}{2} \quad \text{when} \quad \epsilon=\pm\alpha, \nonumber \\
                &=& 0 \quad \; \text{otherwise}, 
    \end{eqnarray}
    for $\alpha > 0$.
   This distribution has a standard deviation $\sigma_{D_2}=\alpha$, which provides the strength of this disorder.\\
   \\
\textbf{\textbullet $\;\;\mathbf{G_1}$: \textit{a modulated Gaussian disorder.}} This is a continuous distribution with mean zero. %whose domain extends from $-\infty$ to $\infty$ and is delineated by the functional form 
    The corresponding probability distribution function is given by
    \begin{equation}
    \label{func1}
        P(\epsilon) = \frac{1}{A_1}|\epsilon^2-1|e^{-(\epsilon^2/a^2)} \;\;\;\;\; -\infty \le \epsilon \le \infty,
    \end{equation}
    where 
    \begin{equation*}
        A_1 = \int_{-\infty}^{\infty} |\epsilon^2-1|e^{-(\epsilon^2/a^2)} d\epsilon,
    \end{equation*}
    and $a>0$.
    The strength of this disorder is characterized by the standard deviation of the distribution,
    \begin{eqnarray}
       \sigma_{G_1} &=& \frac{1}{2} \Big[a^3 \Big(2 \sqrt{\pi } (2-3 a^2)\, \text{erf}\big(\frac{1}{a}\big) \nonumber \\
                &+& a e^{-\frac{1}{a^2}} \Big(\sqrt{\pi } \sqrt{\frac{1}{a^2}} e^{\frac{1}{a^2}} (3 a^2-2)+12\Big)\Big) \Big]^{1/2},\;\;\;
    \end{eqnarray}
    where $\text{erf}(z)=\frac{2}{\sqrt{\pi}}\int_0^z e^{-y^2} dy.$\\
    \\
\textbf{\textbullet$\;\;\mathbf{G_2}$: \textit{a modulated Gaussian disorder.}} Another  continuous distribution with mean zero and a steeper gradient, in comparison to $G_1$, is represented by the distribution function,
     \begin{equation}
     \label{func2}
        P(\epsilon) = \frac{1}{A_2}|\epsilon^2-1|^2e^{-(\epsilon^2/a^2)} \;\;\;\;\; -\infty \le \epsilon \le \infty,
    \end{equation}
     where 
    \begin{equation*}
        A_2 = \int_{-\infty}^{\infty} |\epsilon^2-1|^2 e^{-(\epsilon^2/a^2)} d\epsilon,
    \end{equation*}
    and $a>0$.
    %To assess the "strength" of the disorder introduced, we will refer to the standard deviation of the distribution function from which the disorder parameter is chosen. 
    The standard deviation in this case is 
    \begin{equation}
        \sigma_{G_2} = \Big[\frac{\sqrt{\pi }a^3}{8}  \left(15 a^4-12 a^2+4\right)\Big]^{1/2}.
    \end{equation}\\
    \\
\textbf{\textbullet$\;\;\mathbf{G_3}$: \textit{a modulated Gaussian disorder.}} This represents a distribution sharper than $G_2$. The disorder parameter $\epsilon$ is, in this case, distributed as
     \begin{equation}
     \label{func3}
        P(\epsilon) = \frac{1}{A_3}|\epsilon^2-1|^4e^{-(\epsilon^2/a^2)} \;\;\;\;\; -\infty \le \epsilon \le \infty,
    \end{equation}
     where 
    \begin{equation*}
        A_3 = \int_{-\infty}^{\infty} |\epsilon^2-1|^4 e^{-(\epsilon^2/a^2)} d\epsilon,
    \end{equation*}
    and $a>0$.
    Again we define the strength of the disorder as the standard deviation of the probability distribution, given by
    \begin{equation}
        \sigma_{G_3} = \Big[ \frac{\sqrt{\pi }}{32 a^3} \left(945 a^8-840 a^6+360 a^4-96 a^2+16\right) \Big]^{1/2}.
    \end{equation}

These disorder models provide advantages in the precision of metrological estimation of the relevant quantities, over the same for the ideal scenario with uncorrelated initial probes, in case of evolution by the disordered Hamiltonian $H_1$. 
For uncorrelated initial probes, the disorder-averaged error in frequency estimation is non-existent for the Gaussian (G) and Cauchy-Lorentz (CL) distributions. The median, however exists in these cases, and the disorder-averaged median is found to be beneficial over that of the disorder-free scenario.
Further, the Gaussian and uniform disorder models 
%\textcolor{blue}{\sout{are found to be non-beneficial in this case, but they}}
provide significant enhancement in measurement precision by incorporating some external fields and interaction between the system particles, when copies of two-qubit states are taken as initial probes. The uniform distribution function is also given below, for completeness. \\
\\
%It is worth noting that in these three cases, we have considered a continuous probability distribution with its range extending from $-\infty < \epsilon < \infty$ unlike the discrete scenario.
%\begin{figure}
%\includegraphics[width=8cm]{function_new.pdf}
%\caption{Red - Eq.(~\ref{func1}); green - Eq.(~\ref{func2}); $a=2.0$}
%\label{ent_bas}
%\end{figure}
\textbf{\textbullet $\mathbf{\;\;U}$: \textit{Uniform disorder.}} In this circumstance, $\epsilon$ is distributed as
    \begin{eqnarray}
  \label{uniform}
    P(\epsilon) &=& \frac{1}{s}, \;\;\;\; \text{when} \;\;\;\; -\frac{s}{2} \le \epsilon \le \frac{s}{2} \nonumber \\
                    &=& 0 \;\;\;\;\;\;\;\; \text{otherwise}.
    \end{eqnarray}
    The strength of the disorder, quantified by the standard deviation of the distribution, is given by $\sigma_U = \frac{s}{2\sqrt{3}}$.
   
\subsection{Disorder averaging}
When the disorder parameters of a system are glassy, 
the relevant physical quantities are the glassy disorder-averaged versions of the corresponding quantities, which we will simply refer to as disorder-averaged quantities. For glassy disorder, 
%A proper averaging over the disorder yields practically significant findings of  a disordered physical system. If the disorder parameters are quench disordered, 
the averaging must be done after the relevant physical quantity has been already computed.
%all operations on the system have been performed. 
Suppose that we want to compute the Fisher information-based lower bound of the minimum uncertainty in measuring frequency, in presence of glassy disorder in the system. Then, we first need to evaluate the uncertainty in measuring frequency, $\Delta \tilde{\omega}_{\epsilon}(t)$, encoded in a physical system, for an arbitrary but fixed time $t$, and for an arbitrary but fixed configuration of the disorder. The disorder-averaged deviation in frequency is then given by 
   \begin{equation}
       \Delta \tilde{\omega}_{\mathcal{Q}}(t)=\int_{-\infty}^{\infty}  \Delta \tilde{\omega}_{\epsilon}(t) P(\epsilon) d\epsilon,.
   \end{equation}
   with \(P(\epsilon)\) being the probability distribution of the disorder parameter \(\epsilon\).
  A similar averaging is performed for calculating physically relevant quantities like magnetization and entanglement in glassy systems.
   For discrete probability distributions, the integral is substituted by a summation.
  % Similarly for estimating the coupling constant $J$, the quench averaged error in $J$ is given as
%   \begin{equation}
 %      \int_{-\infty}^{\infty}  \Delta J P(\epsilon) d\epsilon.
 %  \end{equation}
   
  % If this type of integral or summation cannot be addressed analytically, the averaging has to be handled by a numerical method as follows. We have to generate $N$ instances of the disorder $\epsilon$ Haar-uniformly, and let us assume that the instances are denoted as $\epsilon_i$. So, we have $\Delta \tilde{\omega}_{\epsilon_i}$ for each instance.
   %The quantity $\Delta\tilde{\omega}$ is obtained for $N$ realizations of the disorder parameter $\epsilon$, by generating $N$ instances Haar-uniformly and we 
 %  Then we have to take the average of these $N$ instances. This $N$ should be large enough to reach the convergence with a certain precision. The disorder-averaged uncertainty in frequency estimation will take the form
   %Typically $N$ occurrences of the disorder are generated Haar uniformly and if they are designated to as $\epsilon_i$ the quench averaged frequency uncertainty will take the form
%   \begin{equation}
 %     \Delta \tilde{\omega}_{\mathcal{Q}}(t)=\frac{1}{N} \sum_i^N \Delta\tilde{\omega}_{\epsilon_i}(t).
 %  \end{equation}
 %with $N$ being high enough such that convergence is achieved until a certain precision with respect to $N$ is obtained.
%This approach of determining the averaged quantities assumes that the integrals and summations converge to finite values.

 The mean of a probability distribution gives a measure of its central tendency. The mean, however, is non-existent in certain cases. In those cases, or otherwise, it may be useful to use the median, which is the middlemost value of the distribution, and it exists for all probability distributions. If we arrange the functional values of a distribution in increasing (equivalently decreasing) order, and move from the lowest (or highest) value towards the middle, then the middlemost point where one arrives is the median. Since our relevant function is $1/|1+\epsilon|$, which has its maximum at $\epsilon=-1$, we move from the maximum functional value towards the middle. Therefore, to obtain the median, \(x_{M_\epsilon}\), in this case,  the following equation can be solved:
\begin{equation}
\label{median}
    \int_{-1-x_{M_{\epsilon}}}^{-1+x_{M_{\epsilon}}} P(\epsilon) d\epsilon =\frac{1}{2}.
\end{equation}
The solution gives $x_{M_{\epsilon}}$ corresponding to the probability distribution $P(\epsilon)$, from which one can calculate the median $M_{\epsilon}=1/|1+x_{M_{\epsilon}}|$.  In a similar fashion, the first and third quartiles, \(Q_1\) and \(Q_3\) respectively, can be obtained by first solving
%identified as
\begin{equation}
\label{Q}
    \int_{-1-x_{Q_1}}^{-1+x_{Q_1}} P(\epsilon) d\epsilon =\frac{1}{4} \;\;\;\; \text{and} \;\;\;\; \int_{-1-x_{Q_3}}^{-1+x_{Q_3}} P(\epsilon) d\epsilon =\frac{3}{4},
\end{equation}
for \(x_{Q_1}\) and \(x_{Q_3}\) respectively, and then using  $Q_i=1/|1+x_{Q_i}|$, for $i=1,3$. The semi-interquartile range is then given by  $Q=(Q_3-Q_1)/2$.
%%%%%%%%%%%%%%%%%%%%%%%%%%%%

In presence of disorder, a system naturally undergoes disorder-induced fluctuations. As a result, there may occur oscillations in the variance, \(\Delta\), of the estimated parameter, about a mean value with a standard deviation. 
%, which is a measure of dispersion. For completeness, we give a discussion of median in the Supplementary~\cite{median1,median2}. 
Now, if the ``advantage'', $\beta=\Delta \tilde{\omega}_0(t)-\Delta \tilde{\omega}_{\mathcal{Q}}(t)$, of the disordered model over the ideal scenario in a parameter estimation protocol is greater than the standard deviation, $\tilde{\sigma}$, in $\Delta \tilde{\omega}_{\epsilon}(t)$, due to the disorder-induced distribution of the variance of the estimated parameter, then the situation can be referred to as a disorder-induced enhancement over the ideal situation. 
In cases when the mean and standard deviation of \(\Delta\) do not exist, 
%The mean - a measure of central tendency - is non-existent in some cases, and it is then constructive 
we use the median~\cite{median1,median2} and the semi-interquartile range of \(\Delta\).
The semi-interquartile range, like the standard deviation, is a measure of dispersion of a data~\cite{median1,median2}.
%
%We compare the advantage with the corresponding semi-interquartile range, $Q_{\epsilon}$, where we measure the advantage in terms of the median, $M_{\epsilon}$, where $\epsilon$ denotes the disorder parameter. 
%For completeness, we present a short discussion on semi-interquartile range in the SM. 
Here, the disorder-averaged deviation in frequency is given by $\Delta \tilde{\omega}_{\mathcal{Q}}(t)$, $\Delta \tilde{\omega}_0(t)$ represents the Fisher information-based lower bound of minimum error in estimating $\omega$ in an ideal scenario and $\Delta \tilde{\omega}_{\epsilon}(t)$ is the uncertainty in measuring frequency, encoded in a physical system, for an arbitrary but fixed time $t$ and for an arbitrary but fixed configuration of the disorder. In this paper, in all cases, we consider the relative advantage and the corresponding standard deviation, and rechristen them as 
%divide both the quantities 
$\beta$ and 
$\tilde{\sigma}$. This is attained by dividing the original quantities by $\Delta \tilde{\omega}_0(t)$. In particular, this has the effect of making the derived quantities dimensionless.
Suffixes of the quantities will be used to refer to the different disorder distributions used.

We begin by finding the disorder-averaged minimum uncertainty in frequency $\omega$, in presence of the disorder parameter $\epsilon$, incorporated in the Hamiltonian $H_1$. For a single-qubit system, with the encoding being effected by a unitary evolution with the unitary $U_1=e^{-\frac{i}{\hbar}H_1 t}$, the optimal input state for measuring $\omega$ is $\ket{\psi_0}=(\ket{0}+\ket{1})/\sqrt{2}$, where $\ket{0}$ and $\ket{1}$ are eigenvectors of $H_1$. 
%following the prescription in Eq.~(\ref{best_input}) and 
%\textcolor{red}{\sout{The QFI evaluated using Eq.~(11) of the Supplementary material, in this case is found to be $F_{Q}^\epsilon=t^2(1+\epsilon)^2$.}}
The QFI in this case is 
%determined to be 
$F_{Q}^\epsilon=t^2(1+\epsilon)^2$.
%, and the proof of this result is presented in the SM.
As Fisher information is additive, we can extend this situation to $\nu$ copies of $\ket{\psi_0}$, i.e., $\ket{\psi_0}^{\otimes \nu}$, and the disorder-averaged minimum deviation in the estimation of the frequency is given by
\begin{equation}
    \Delta \tilde{\omega}^{(p)}_{\mathcal{Q}}(t)=\int_{\text{lb}}^{\text{ub}}\frac{1}{|1+\epsilon|t\sqrt{\nu}} \; P(\epsilon) d\epsilon,
\end{equation}
where $\text{lb}$ and $\text{ub}$ denote respectively the lower bound and upper bound of the corresponding probability distribution function from which $\epsilon$ is chosen. We compare the disorder-averaged quantities for the different probability distributions, with the ideal minimum uncertainty in $\omega$, which is $\Delta \tilde{\omega}^{(p)}_0=\frac{1}{t\sqrt{\nu}}$. \\

%%%%%%%%%%%%%%%%%%%%%%%%%%%%%%%%%%%%%%%%%
\section{%Measurement precision in presence of glassy disorder for uncorrelated initial probes
Unbiased estimators and the quantum Cram\'{e}r-Rao bound in presence of glassy disorder}
\label{extra}
We consider the insertion of a disorder parameter, into a single-qubit system described by the Hamiltonian 
%$\mathcal{H}_1$. 
\( \mathcal{H}_1=-\hbar\omega \ket{1}\langle 1|\), with \(\omega\) being the parameter to be estimated.
We incorporate the disorder in two ways into the ``ideal" Hamiltonian. The altered forms of the Hamiltonian, with disorder, are taken as
\begin{equation}
\label{ham}
   H_1 = -\hbar\omega(1+\epsilon)\ketbra{1}{1} 
  \end{equation}
  %\\
%   &&\text{and} \quad H_2 = -\hbar(\omega+\epsilon)\ketbra{1}{1}, 
%\end{eqnarray}
and 
\(H_2 = -\hbar(\omega+\epsilon)\ketbra{1}{1}\),
where the disorder is modeled by the parameter $\epsilon$.
An unbiased estimator for $\omega$ can be found in this disordered situation, and the Cram\'er-Rao bound therein can also be derived. 
An explicit derivation is 
%A detailed discussion on the unbiased estimator of $\omega$, and the derivation of the Cram\'er-Rao bound in these disordered situations, is 
provided below.
%in the succeeding subsection.
%in the SM.
\\

%%%%%%%%%%%%%%%%%%%%%%%%%%%%%%%%%%%%%%%%%%%%%%%%%%%%%%%%%%%%%%%%%%%
%\subsection{Unbiased estimators and the quantum Cram\'{e}r-Rao bound in presence of glassy disorder}
%\label{unbiased}
For a single-qubit system, with the encoding being effected by a unitary evolution, 
%with the unitary 
$U_1=e^{-\frac{i}{\hbar}H_1 t}$, the optimal input state for measuring $\omega$ is $\ket{\psi_{0}}=(\ket{0}+\ket{1})/\sqrt{2}$, with $\ket{0}$ and $\ket{1}$ being the eigenvectors of $H_1$.
%(Eq.~(\ref{eijey})). 
We perform measurements with projectors from the eigenbasis of SLD of the encoded state, $\ket{\psi_{\omega}}=(\ket{0}+e^{i\omega t}\ket{1})/\sqrt{2}$. Let these measurement operators belong to the set, $\{M,I_2-M\}$. 
Suppose that for the measurement 
%we measure using the 
operator $M$, the corresponding eigenvalue is $x_1$ and it clicks with probability $p_1=\text{Tr}(M\ketbra{\psi_{\omega}}{\psi_{\omega}})$, while for $I_2-M$, the eigenvalue is $x_2$ and clicks with probability $p_2=\text{Tr}((I_2-M)\ketbra{\psi_{\omega}}{\psi_{\omega}})$.  The explicit forms of these probabilities are $p_1=[1+\cos(\omega t(1+\epsilon))]/2$, and $p_2=[1-\cos(\omega t(1+\epsilon))]/2$. 
%The measurement outcomes in this situation, $x_1$ and $x_2$, occur with probabilities $p_1$ and $p_2$ respectively. 
From these outcomes, we have to estimate the unknown parameter, $\omega$. The time of the measurement 
%experiment is given by 
is $t$ (where the initial time is \(0\)), which we assume to know exactly, and the disorder parameter $\epsilon$, being glassy, is known to us, although can take arbitrary values within a given distribution that depends on the material used. The disorder parameter, therefore, varies unpredictably from one trial to another, but once realized, it is known and fixed for all measurements performed in that trial, and is distributed according to a known probability density.
%\\
%\\
A suitable estimator in our case is \(\hat{\omega}(x):\{x_1, x_2\} \to \mathbb{R}^+\cup \{0\}\), given by 
\begin{eqnarray}
\label{est}
    \hat{\omega}(x_1)=\frac{\cos^{-1}(2p_1-1)}{t(1+\epsilon)}, \;\;\;\;
    \hat{\omega}(x_2)=\frac{\cos^{-1}(1-2p_2)}{t(1+\epsilon)}. \;\;\;
\end{eqnarray}
%The domain of the function comprises of the points, $x_1$ and $x_2$, which are mapped respectively to $\hat{\omega}(x_1)$ and $\hat{\omega}(x_2)$. 
The values of $t$ and $\epsilon$ are pre-determined or determined from an analysis of the glassy material, and $p_1$ and $p_2 (= 1-p_1)$ are obtained after the measurements, and can be utilized in Eq.~\eqref{est} to estimate the unknown parameter, $\omega$. 
It may be noted that 
%A peculiarity of the problem in this scenario is that, 
for the scenario at hand, the estimators are not explicitly dependent on the values of the random variables, $x_1$ and $x_2$. We now try to see whether this estimator is unbiased. The average of the estimator function is given by
\begin{eqnarray}
\label{av}
    \langle \hat{\omega} (x) \rangle = \hat{\omega}(x_1) p_1 + \hat{\omega}(x_2)p_2.
\end{eqnarray}
If we substitute the expressions of $p_1$ and $p_2$ in Eq.~\eqref{av}, then we obtain, $\langle \hat{\omega} (x) \rangle = \omega (p_1+p_2)=\omega$. This proves that the estimator function is indeed unbiased.

We now derive the quantum Cram\'{e}r-Rao bound involving the standard deviation of the estimator, for the disordered scenario. We begin with the identity given in Eq.~\eqref{av}, which can be rewritten as
\begin{eqnarray}
\label{av1}
    \sum_{i=1}^2 \Delta \omega_i p_i = 0,
    \end{eqnarray}
where $\Delta\omega_i= \hat{\omega}(x_i) - \langle \hat{\omega} \rangle$. Now if we take the derivative of the left hand side of Eq.~\eqref{av1}, we obtain
\begin{eqnarray}
\label{dif}
    \sum_{i=1}^2 \Delta \omega_i (p_i \frac{\partial \ln p_i}{\partial \omega}) &=& -\sum_{i=1}^2 \frac{\partial  (\Delta \omega_i)}{\partial \omega} p_i \nonumber \\
    &=&  \frac{\partial \langle\hat{\omega}\rangle}{\partial \omega}.
\end{eqnarray}
The second equality follows from the fact that $\hat{\omega}(x_i)$ is independent of $\omega$. Now, an use of the Cauchy-Schwartz inequality in Eq.~\eqref{dif} yields 
\begin{eqnarray}
\label{cr_in}
    \left(\sum_{i=1}^2 \Delta \omega_i \right)^2 \left(\sum_{i=1}^2 p_i \frac{\partial \ln p_i}{\partial \omega} \right)^2 \ge \left|\frac{\partial \langle\hat{\omega}\rangle}{\partial \omega} \right|^2 && \nonumber \\
    \implies \left(\sum_{i=1}^2 \Delta \omega_i \right)^2 \ge \left(\sum_{i=1}^2 p_i \frac{\partial \ln p_i}{\partial \omega} \right)^{-2}, &&
\end{eqnarray}
which is  the Cram\'{e}r-Rao bound. The second inequality~\eqref{cr_in} follows from the unbiased nature of the estimator, i.e., $\langle \hat{\omega} (x) \rangle = \omega$ in our case. Finally, if we evaluate the quantity on the right hand side of inequality~\eqref{cr_in}, using the expressions of $p_1$ and $p_2$, we get
\begin{equation}
\label{final}
    \left(\sum_{i=1}^2 \Delta \omega_i \right)^2 \ge \frac{1}{t^2(1+\epsilon)^2},
\end{equation}
where $t$ denotes the time duration for which the unitary operator, $U_1$, acts on the probe, and $\epsilon$ is the disorder parameter. The parameter \(t\) is also the time at which the SLD measurement is performed.
Since this bound (Eq.~\eqref{final}) corresponds to an optimal measurement, it actually gives the QFI, and the bound is the quantum Cram{\'e}r-Rao inequality in this case.
Therefore, the QFI, in this scenario, is given by $F_Q^{\epsilon}=t^2(1+\epsilon)^2$, for a single copy of the probe.
\\
%%%%%%%%%%%%%%%%%%%%%%%%%%%%%%%%%%%%%%%%%%%%%%%%%%%%%%%%%%%%%%%%%%%

%%%%%%%%%%%%%%%%%%%%%%%%%%%%%%%%%%%%%%%
%%%%%%%%%%%%%%%%%%%%%%%%%%%%
\section{%Response to Cauchy-Lorentz, discrete and modulated Gaussian distributions
Response to glassy disorder distributions in estimating frequency}
\label{Sec:5}
In this section, we present specific instances of ``order from disorder" scenarios, which can be viewed as a valuable aspect in the design of a highly precise atomic clock. We consider the disordered Hamiltonians $H_1$ and $H_2$ (see Eq.~(\ref{ham})), where the disorder parameter $\epsilon$ is selected randomly from several paradigmatic distributions, both discrete and continuous. While %\textcolor{blue}{\sout{In this paper, while}} 
considering single-qubit systems, we have utilized the Gaussian ($G$), Cauchy-Lorentz ($CL$), two discrete ($D_1$ and $D_2$) and three modulated Gaussian  ($G_1$, $G_2$ and $G_3$) probability distributions for uncorrelated initial probes.
%We consider the probability distribution functions, some of which are discrete and some are continuous, and investigate whether they can offer a disorder-induced enhancement in estimation of a parameter over their ideal scenario. 
%Among a few disorder models of different probability distributions, 
We will find that the metrological estimation provides order from disorder for certain cases. 
%%%%%%%%%%%%%%%%%%%%%%%%%%
\subsection{Gaussian glassy disorder $\mathbf{G}$} 
\begin{figure}
\includegraphics[width=9cm]{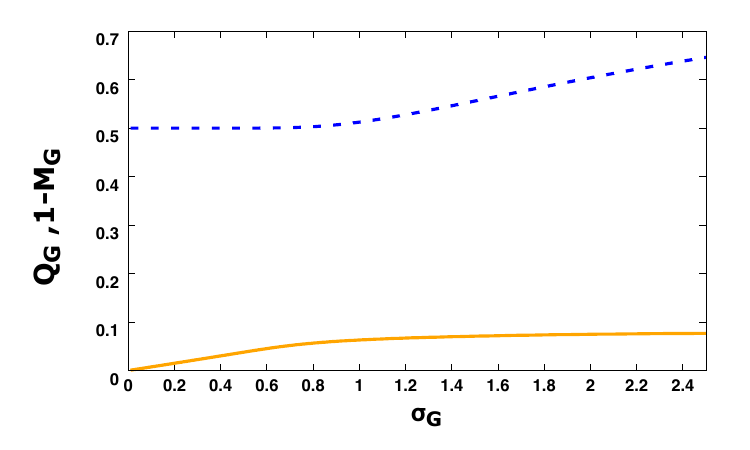}
\caption{Disorder-induced enhancement in frequency estimation for uncorrelated initial probes. Here we depict the difference between the true value of the median with the disorder-averaged median in the estimation of $\omega$ (blue dashed curve), and compare with its semi-interquartile range (orange curve). The subscript, $G$, denotes that the disorder parameter $\epsilon$ is chosen randomly from a Gaussian distribution $G$. 
%\textcolor{blue}{\sout{in (a) and from $G_2$ in (b). The blue  surface (dashed line) in panel (a) ((b)) depicts the quantity $\beta_{D_2}$ ($\beta_{G_2}$) and the corresponding yellow surface (solid line) presents the standard deviation $\tilde{\sigma}_{D_2}$ ($\tilde{\sigma}_{G_2}$).}} 
All quantities plotted are dimensionless. 
%(b) The yellow curve depicts the standard deviation in $\Delta\tilde{\omega}_1/\Delta\tilde{\omega}_0$, given by $\sigma'$, and the blue curve gives the ratio $\beta=\frac{\Delta\tilde{\omega}_0-\Delta\tilde{\omega}_1}{\Delta\tilde{\omega}_0}$ along the y-axis. All quantities plotted are dimensionless.
}
\label{gauss_med}
\end{figure}

\begin{figure}
\includegraphics[width=7.5cm]{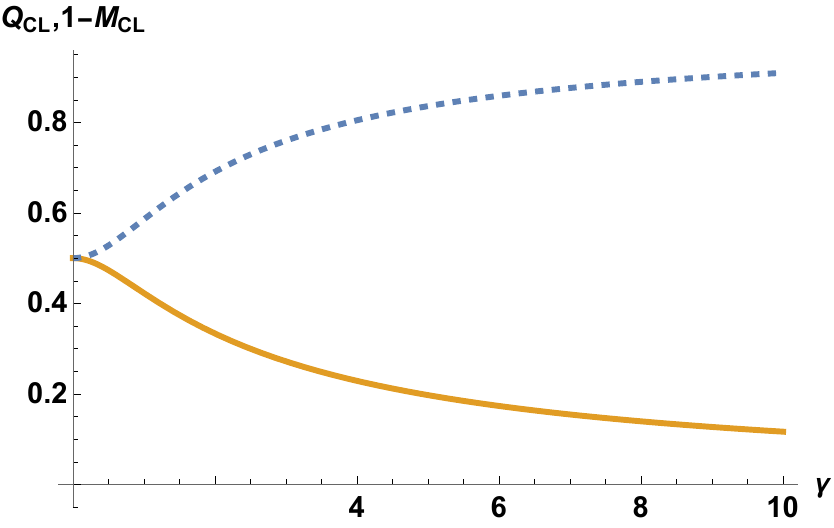}
\caption{Disorder-induced enhancement in frequency estimation for uncorrelated initial probes. Here we depict the difference between the true value of the median and the disorder-averaged median in the estimation of $\omega$ (blue dashed curve), and compare with its semi-interquartile range (orange curve). The subscript, $CL$, denotes that the disorder parameter $\epsilon$ is chosen randomly from a Cauchy-Lorentz distribution. 
All quantities plotted are dimensionless.}
\label{CL_med}
\end{figure}

 We calculate the disorder-averaged median, $M_G=1/|1+x_{M_{G}}|$, by solving $x_{M_{G}}$ from
 $\tilde{\text{erf}}(x_{M_{G}})=(\text{erf}(1+x_{M_{G}})- \text{erf}(1-x_{M_{G}}))/ \sqrt{2} \sigma_G$
 %$x_{M_{G}}=  \tilde{\text{erf}}^{-1}\big[(\text{erf}(1+x_{M_{G}})- \text{erf}(1-x_{M_{G}}))/ \sqrt{2} \sigma_G\big]$,
  corresponding to the probability distribution $G$. Note that the mean does not exist in this case.
%  , and so we calculate the median.  
Fig.~\ref{gauss_med} depicts that the advantage, $1-M_G$, is positive and is significantly higher than the semi-interquartile range, $Q_G$, with the standard deviation, $\sigma_G$, of $G$  plotted along horizontal axis. Thus the presence of disorder is advantageous over the disorder-free scenario, hence displaying order from disorder.

\begin{figure*}
\includegraphics[width=\textwidth]{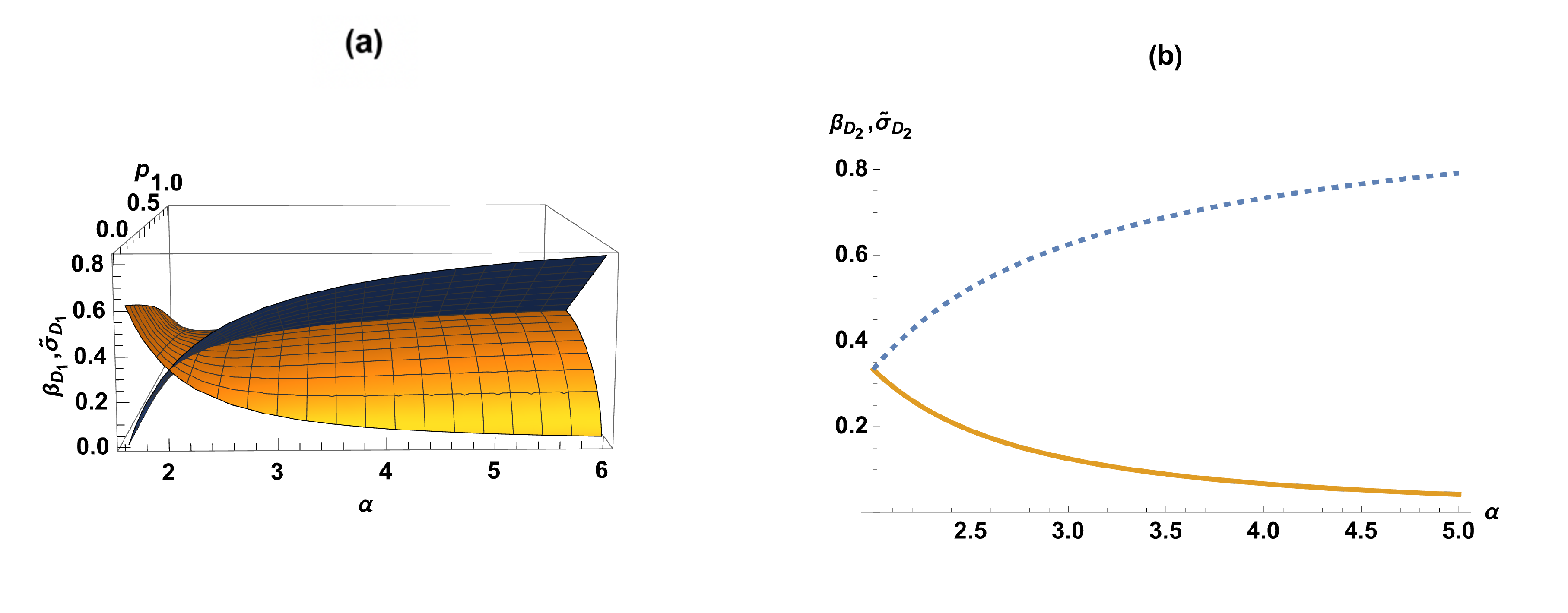}
\caption{Disorder-induced enhancement in frequency estimation for uncorrelated initial probes. Here we depict the effect of discrete glassy disorder on the minimum deviation of estimation of $\omega$, encoded in each of the initial states by evolving them with the Hamiltonian $H_1$. The disorder parameter $\epsilon$ is chosen randomly from the discrete distribution $D_1$ in (a) and from $D_2$ in (b). The blue  surface (dashed line) in panel (a) ((b)) depicts the quantity $\beta_{D_1}$ ($\beta_{D_2}$) and the corresponding yellow surface (solid line) presents the standard deviation $\tilde{\sigma}_{D_1}$ ($\tilde{\sigma}_{D_2}$). All qunatities plotted are dimensionless. 
%(b) The yellow curve depicts the standard deviation in $\Delta\tilde{\omega}_1/\Delta\tilde{\omega}_0$, given by $\sigma'$, and the blue curve gives the ratio $\beta=\frac{\Delta\tilde{\omega}_0-\Delta\tilde{\omega}_1}{\Delta\tilde{\omega}_0}$ along the y-axis. All quantities plotted are dimensionless.
}
\label{1q_discrete1}
\end{figure*}

%\vspace{0.2cm}
%\noindent 
\subsection{Cauchy-Lorentz glassy disorder $\mathbf{CL}$} In this case, the disorder-averaged median, $M_{CL} = 1/|1+x_{M_{CL}}|$ , can be evaluated by solving~\cite{footnote} 
%$x_{M_{CL}}$ from
\begin{equation}
    \frac{1}{\pi}\Big[\tan^{-1}\Big(\frac{x_{M_{CL}}+1}{\gamma}\Big) + \tan^{-1}\Big(\frac{x_{M_{CL}}-1}{\gamma}\Big) \Big]=\frac{1}{2},
\end{equation}
which gives $x_{M_{CL}}=\sqrt{1+\gamma^2}$.
%~\cite{root}. 
Note that the mean does not  exist in this case also, and hence we have looked at the median.
%The disorder-averaged median in this case is $M_{CL}=1/(1+\sqrt{1+\gamma^2})$, and 
The relative advantage,  $1-M_{CL}$, is always positive. We have also compared this advantage with the semi-interquartile range, which, in this case, is given by $Q_{CL}=(1+\sqrt{1+2\gamma^2})/((1+\sqrt{1+2\gamma^2})^2-\gamma^2)$, which is less than the relative advantage, $1-M_{CL}$, in a certain range of $\gamma$ (refer to Fig.~\ref{CL_med}).
Thus we again obtain the phenomenon of  order from disorder.

%%%%%%%%%%%%%%%%%%%%%%%%%%
%\textcolor{magenta}{\subsection{Cauchy-Lorentz distribution}
%The discussion about the response to Cauchy-Lorentz distribution on the estimation of frequency \(\omega\) of the Hamiltonian in Eq.~(\ref{eijey}) is presented in the main text. Here we provide an illustration of that result in Fig.~\ref{CL_med}.}

\subsection{Discrete glassy disorder of type $\mathbf{D_1}$} For the disorder parameter $\epsilon$ chosen from the discrete distribution, $D_1$, the disorder-averaged frequency deviation, for $\nu$ copies of uncorrelated probes, is given by
\begin{equation}
     \Delta \tilde{\omega}^{(p)}_{\mathcal{Q}_{D_1}}(t) = \frac{1}{t\sqrt{\nu}}\Big[\frac{1-p}{2}\Big(\frac{1}{|1-\alpha|}+\frac{1}{1+\alpha}\Big)+p \Big].
\end{equation}
At $\alpha=1$, the function $\Delta \tilde{\omega}^{(p)}_{\mathcal{Q}_{D_1}}(t)$ diverges. 
%We now consider two scenarios, $\alpha >1$ and $\alpha<1$.  %\textcolor{red}{The condition, $\alpha>1$ is physically meaningful if $p$ is sufficiently close to unity.}
%while the second is when $\alpha<1$. Corresponding to the first case, 
We find that for $\alpha \ge \frac{1+\sqrt{5}}{2} \approx %\frac{1}{2}+\frac{\sqrt{5}}{2} \approx 
1.618$, the error in frequency in presence of glassy disorder of type $D_1$, has an advantage over the ideal scenario i.e., $\Delta \tilde{\omega}_{\mathcal{Q}_{D_1}}^{(p)}(t)<\Delta \tilde{\omega}^{(p)}_0(t)$, $\forall t$. 
%In this  the error in frequency in presence of disorder has an advantage over the disorder free scenario. 
The blue surface of Fig.~\ref{1q_discrete1}-(a) presents the relative advantage, $\beta_{D_1}=\frac{\Delta \tilde{\omega}^{(p)}_0(t)-\Delta \tilde{\omega}_{\mathcal{Q}_{D_1}}^{(p)}(t)}{\Delta \tilde{\omega}^{(p)}_0(t)}$, with respect to $\alpha$ and $p$. 
%\textcolor{red}{The figure corresponding to this case is given in Supplementary.
%The blue surface in Fig.1a in Supplementary} presents the relative advantage, $\beta_{D_1}=\frac{\Delta \tilde{\omega}^{(p)}_0(t)-\Delta \tilde{\omega}_{\mathcal{Q}_{D_1}}^{(p)}(t)}{\Delta \tilde{\omega}^{(p)}_0(t)}$, with respect to $\alpha$ and $p$. 
%We consider the difference between the mean values of the ordered and disordered frequency errors, which is made dimensionless by dividing by $\Delta\tilde{\omega}_0$, and denote this quantity as $\beta=\frac{\Delta\tilde{\omega}_0-\Delta\tilde{\omega}_1}{\Delta\tilde{\omega}_0}$.
$\beta_{D_1}$ exhibits positive values, which signifies a possibility of obtaining an order from disorder situation for the discrete glassy disorder $D_1$. But, for further assurance, we have to compare $\beta_{D_1}$ with the standard deviation in $\Delta \tilde{\omega}_{\epsilon_{D_1}}^{(p)}(t)/\Delta \tilde{\omega}_0^{(p)}(t)$. Suppose, the standard deviation is denoted as $\tilde{\sigma}_{D_1}$.
%is plotted with respect to $\alpha$ along x-axis and time $t$ along y. 
The behavior of $\tilde{\sigma}_{D_1}$ is shown by  the yellow surface of the said figure, which 
%It is emphasised that the fluctuations in $\Delta \tilde{\omega}_1/\Delta \tilde{\omega}_0$, depicted by its standard deviation $\sigma'$, are 
is much smaller than $\beta_{D_1}$ across a wide range of the $\alpha-p$ space for $\alpha  \gtrsim 
1.618$. This implies that we can have disorder-induced enhancements in efficiency of the frequency measurement for $\alpha  \gtrsim 
1.618$ for the disorder model $D_1$. 

%For $\alpha < 1$, there is no advantage on introduction of a quenched disorder parameter in the system for this type of disorder.

%\vspace{0.2cm}

%\noindent 
\subsection{Discrete glassy disorder of type $\mathbf{D_2}$} If $\epsilon$ is chosen randomly from the discrete distribution $D_2$, the disorder-averaged minimum deviation in estimation of frequency, for $\nu$ copies of uncorrelated probes, is given by
\begin{equation}
    \Delta \tilde{\omega}_{\mathcal{Q}_{D_2}}^{(p)}(t) = \frac{1}{2t\sqrt{\nu}}\Big[\frac{1}{|1-\alpha|}+\frac{1}{1+\alpha} \Big].
\end{equation}
%\begin{figure}
%\includegraphics[width=8cm]{1q_discrete2.pdf}
%\caption{The yellow curve depicts the standard deviation in $\Delta\tilde{\omega}_1/\Delta\tilde{\omega}_0$, given by $\sigma'$, and the blue curve gives the ratio $\beta=\frac{\Delta\tilde{\omega}_0-\Delta\tilde{\omega}_1}{\Delta\tilde{\omega}_0}$ along the y-axis. All quantities plotted are dimensionless.}
%\label{1q_discrete2}
%\end{figure}
Like in the case of $D_1$, here also $\Delta \tilde{\omega}_{\mathcal{Q}_{D_2}}^{(p)}(t)$ diverges for $\alpha=1$ and the advantage, i.e., $\Delta \tilde{\omega}_{\mathcal{Q}_{D_2}}^{(p)}(t)<\Delta \tilde{\omega}^{(p)}_0(t)$, is obtained only for the region, $\alpha \ge \frac{1+\sqrt{5}}{2} \approx 1.618$
%and previously, there may be potentially two situations corresponding to $\alpha<1$ and $\alpha>1$. Nevertheless there is a benefit over the disorder-free scenario only when $\alpha>1$. Precisely, the range of $\alpha$ for which $\Delta \tilde{\omega}_1 < \Delta \tilde{\omega}_0$ is 
%$\alpha 
%> \frac{1}{2}+\frac{\sqrt{5}}{2}
%\gtrsim 1.618$. 
%However, because $\alpha>1$ implies that $\epsilon$ is drawn from a wide range of interval, this result is physically less significant than the type 1 discrete disorder. 
%The $\alpha<1$ situation is not possible for $\Delta \tilde{\omega}_1/\Delta \tilde{\omega}_0<1$.
The behavior of the relative advantage, $\beta_{D_2}=\frac{\Delta \tilde{\omega}^{(p)}_0(t)-\Delta \tilde{\omega}_{\mathcal{Q}_{D_2}}^{(p)}(t)}{\Delta \tilde{\omega}^{(p)}_0(t)}$, along with the standard deviation in $\Delta \tilde{\omega}_{\epsilon_{D_2}}^{(p)}(t)/\Delta \tilde{\omega}_0^{(p)}(t)$, represented by $\tilde{\sigma}_{D_2}$, with respect to $\alpha$ are depicted in Fig.~\ref{1q_discrete1}-(b), for $\alpha \gtrsim 1.618$. It should be noted that $\tilde{\sigma}_{D_2}$ is significantly lower than $\beta_{D_2}$. This suggests that there occurs the order from disorder effect using this distribution for $\alpha \gtrsim 1.618$.

\begin{figure*}
\includegraphics[width=5.9cm, height=4.1cm]{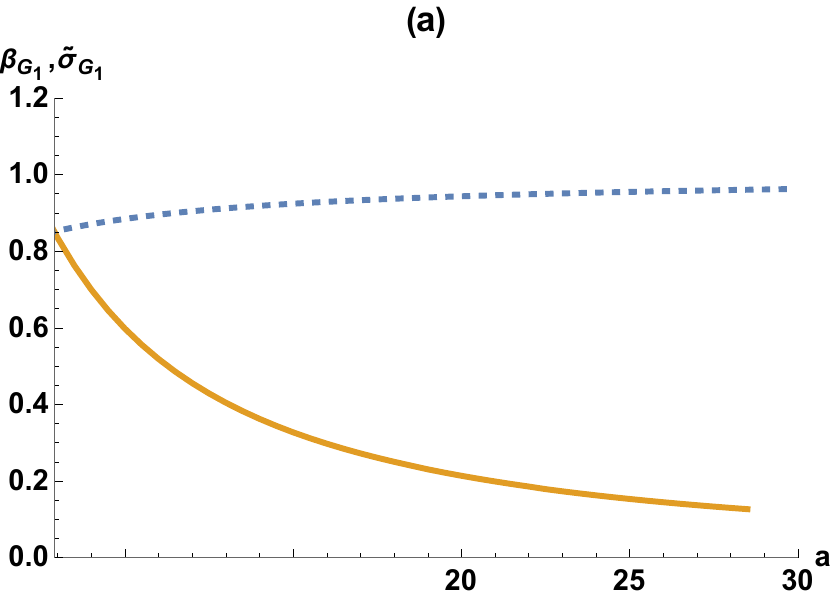}
\includegraphics[width=5.5cm]{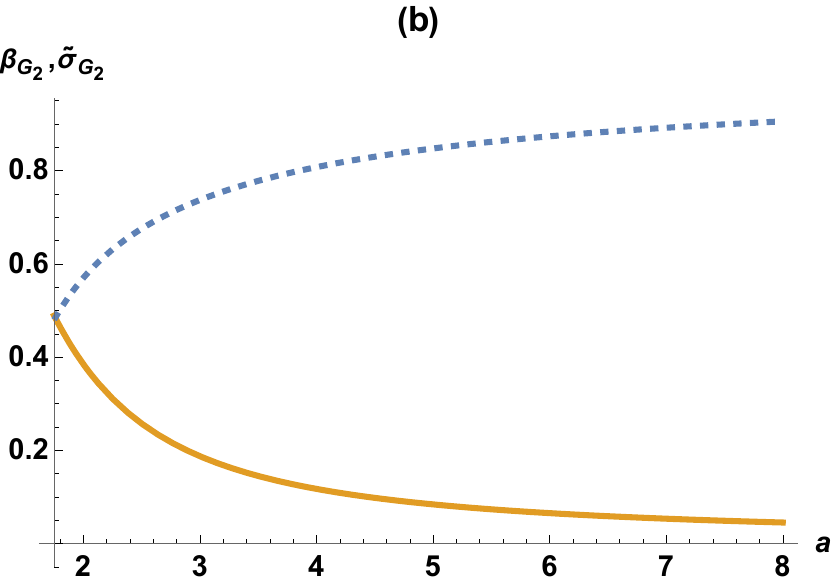}
\includegraphics[width=5.5cm]{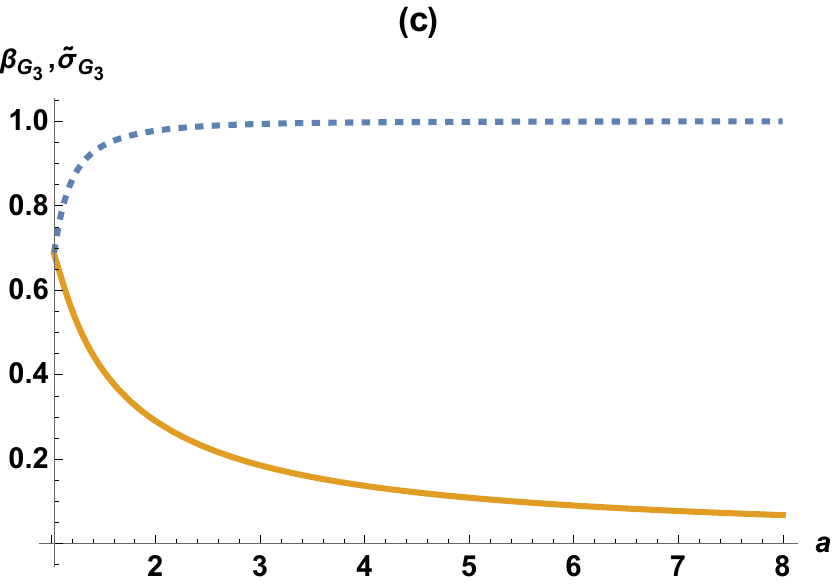}
\caption{Disorder-induced enhancement in frequency estimation for uncorrelated initial probes. Here we present the effect of modulated Gaussian glassy disorders on the minimum deviation of estimation of $\omega$, encoded in each of the initial states by evolving them with the Hamiltonian $H_1$. 
This figure depicts the quantity $\tilde{\beta}_{G_i}=\frac{\Delta \tilde{\omega}^{(p)}_{\mathcal{Q}_{G_i}}(t)}{\Delta\tilde{\omega}_0^{(p)}(t)}$ for $i=1,2,3$, denoting the three modulated Gaussian distributions $G_1$, $G_2$ and $G_3$ respectively. 
The quantities $\{\beta_{G_1},\tilde{\sigma}_{G_1}\}$, $\{\beta_{G_2},\tilde{\sigma}_{G_2}\}$ and $\{\beta_{G_3},\tilde{\sigma}_{G_3}\}$ are plotted with respect to the parameter $a$ in case of the distributions $G_1$, $G_2$ and $G_3$ respectively in panels (a), (b) and (c). The quantities $\beta_{G_i}$ are is presented by the blue dashed lines and the quantities $\tilde{\sigma}_{G_i}$ are demonstrated with yellow solid lines in all the  panels. 
%The ratio of the disorder averaged error in frequency to the frequency error in absence of disorder as a function of the parameter $a$ present in the modified Gaussian functions. There is a considerable range of $a$ where the ratio is less than identity suggesting an advantage in presence of disorder over the disorder free scenario. (b) The yellow curve gives the standard deviation in $\Delta\tilde{\omega}_1/\Delta\tilde{\omega}_0$, given by $\sigma'$, and the blue curve shows the ratio $\beta=\frac{\Delta\tilde{\omega}_0-\Delta\tilde{\omega}_1}{\Delta\tilde{\omega}_0}$ along the y-axis. Here $\beta > 0$ and there is a finite difference between the two curves, suggesting an advantage over the disorder-less case. The quantities plotted along both the axes are dimensionless. (c) The yellow curve depicts the standard deviation in $\Delta\tilde{\omega}_1/\Delta\tilde{\omega}_0$, given by $\sigma'$, and the blue curve gives the ratio $\beta=\frac{\Delta\tilde{\omega}_0-\Delta\tilde{\omega}_1}{\Delta\tilde{\omega}_0}$ along the y-axis. A benefit in presence of disorder is noted. 
All  quantities plotted along the horizontal and vertical axes are dimensionless.}
\label{modulated_gauss}
\end{figure*}

\subsection{Modulated Gaussian glassy disorder of type $\mathbf{G_1}$} For the modulated Gaussian  disorder of type $G_1$, the minimum error in estimation of frequency, \(\omega\), of the Hamiltonian \(H_1\) of Eq.~(\ref{ham}), for $\nu$ uncorrelated initial probes, is
%\begin{widetext}
\begin{equation}
    %\Delta \tilde{\omega}^{(p)}_{\mathcal{Q}_{G_1}} = \frac{1}{t\sqrt{\nu}}\frac{2a+e^{1/a^2}\sqrt{\pi}+2e^{1/a^2}\sqrt{\pi}\;\text{erf}(\frac{1}{a})-1}{\big[-4a+2e^{1/a^2}\sqrt{\pi}-a^2 e^{1/a^2}\sqrt{\pi}-4 e^{1/a^2}\sqrt{\pi}\;\text{erf}(\frac{1}{a})+2a^2 e^{1/a^2}\sqrt{\pi}\;\text{erf}(\frac{1}{a})\big]}.
    \Delta \tilde{\omega}^{(p)}_{\mathcal{Q}_{G_1}} = \frac{1}{t\sqrt{\nu}}\frac{2a+e^{1/a^2}\sqrt{\pi}+2e^{1/a^2}\sqrt{\pi}\;\text{erf}(\frac{1}{a})-1}{\big[-4a + \sqrt{\pi} e^{1/a^2} (2-a^2)(1-2 \text{erf}(\frac{1}{a}))\big]},\;\;\;\;
\end{equation}
where $\text{erf}(z)=\frac{2}{\sqrt{\pi}}\int_0^z e^{-y^2} dy.$
%\end{widetext}
%where $\text{erf}(z)=\frac{2}{\sqrt{\pi}}\int_0^z e^{-t^2} dt$.
%and $\text{erfc}(z)$ denotes the complementary error function given by $\text{erfc}(z)=1-\text{erf}(z)$. 
From the nature of the quantity $\frac{\Delta \tilde{\omega}^{(p)}_{\mathcal{Q}_{G_1}}(t)}{\Delta\tilde{\omega}_0^{(p)}(t)}$, with the variation of $a$ (see Fig.~\ref{modulated_gauss}-(a)), it is evident that this glassy system allows $\Delta \tilde{\omega}_{\mathcal{Q}_{G_1}}^{(p)}(t) < \Delta \tilde{\omega}_0^{(p)}(t)$. For this particular distribution, the standard deviation in $\Delta\tilde{\omega}_{\epsilon_{G_1}}^{(p)}(t)/\Delta \tilde{\omega}_0^{(p)}(t)$, %\textcolor{blue}{\sout{is non-existent.}} 
given by $\tilde{\sigma}_{G_1}$, is also less than the corresponding $\tilde{\beta}_{G_1}$, hence exhibiting order from disorder.

\subsection{Modulated Gaussian glassy disorder of type $\mathbf{G_2}$} For this glassy disorder situation, the disorder parameter, $\epsilon$, is chosen from the distribution $G_2$. The disorder-averaged frequency deviation for this case for $\nu$ copies of initial uncorrelated probes is
\begin{equation}
  \Delta \tilde{\omega}_{\mathcal{Q}_{G_2}}^{(p)} = \frac{1}{t\sqrt{\nu}}\Big[\frac{4 \big[ (a+a^3)\frac{1}{\sqrt{e}}-\frac{1}{2}(a^2-2)\sqrt{\pi}\;\text{erf}(\frac{1}{a}) \big]}{(4-4a^2+3a^4)\sqrt{\pi}}\Big].
\end{equation}
%In Fig.~3 of the Supplementary material, we present the quantity $\frac{\Delta\tilde{\omega}^{(p)}_{\mathcal{Q}_{G_2}}(t)}{\Delta\tilde{\omega}_0^{(p)}(t)}$ with the green dashed line, which shows that this type of glassy disorder provide better precision in a larger region of $a$ than the same of $G_1$.  
% depicts that this function is advantageous over the disorder free situation as there is a certain range of $a$ in which $\Delta \tilde{\omega}_1/\Delta \tilde{\omega}_0<1$.
Fig.~\ref{modulated_gauss}-(b) 
%of the Supplementary material 
shows that the relative advantage, $\beta_{G_2}=\frac{\Delta\tilde{\omega}_0^{(p)}(t)-\Delta \tilde{\omega}^{(p)}_{\mathcal{Q}_{G_2}}(t)}{\Delta\tilde{\omega}_0^{(p)}(t)}$, is much greater than $\tilde{\sigma}_{G_2}$, which is the standard deviation in $\Delta\tilde{\omega}^{(p)}_{\epsilon_{G_2}}(t)/\Delta\tilde{\omega}_0^{(p)}(t)$. So, the occurrence of an order from disorder situation is prominently present in this situation.
%The effects of the distributions $G_1$ and $G_3$ in frequency estimation are given in the Supplementary material. 

\subsection{Modulated Gaussian glassy disorder of type $\mathbf{G_3}$} 
\label{appen1-natun}
%\textbf{Modulated Gaussian glassy disorder of type $\mathbf{G_3}$:} 
%The disorder-averaged frequency deviation for $\nu$ copies of initial uncorrelated probes is given in this case by \(\Delta \tilde{\omega}^{(p)}_{\mathcal{Q}_{G_3}}\), whose analytical form is given in \textcolor{red}{the Supplementary material.}
%ekhane!
%\begin{widetext}
%\begin{equation}
%    \Delta \tilde{\omega}^{(p)}_{\mathcal{Q}_{G_3}} = \frac{1}{t\sqrt{\nu}} \frac{4\big[2a(2-3a^2+4a^4)e^{-1/a^2} + (4-4a^2+3a^4)\sqrt{\pi}\;\text{erf}(\frac{1}{a}) \big]}{(16-32a^2+72a^4-120a^6+105a^8)\sqrt{\pi}}.
%\end{equation}
%\end{widetext}
The disorder-averaged frequency deviation for $\nu$ copies of initial uncorrelated probes is given in this case by 
%\(\Delta \tilde{\omega}^{(p)}_{\mathcal{Q}_{G_3}}\), whose analytical form is given in the Appendix.
\begin{widetext}
\begin{equation}
    \Delta \tilde{\omega}^{(p)}_{\mathcal{Q}_{G_3}} = \frac{1}{t\sqrt{\nu}} \frac{4\big[2a(2-3a^2+4a^4)e^{-1/a^2} + (4-4a^2+3a^4)\sqrt{\pi}\;\text{erf}(\frac{1}{a}) \big]}{(16-32a^2+72a^4-120a^6+105a^8)\sqrt{\pi}},
\end{equation}
\end{widetext}
for estimating the frequency \(\omega\) of the Hamiltonian in Eq.~(\ref{ham}).
Here the range of the distribution parameter $a$, providing an enhancement in frequency estimation, 
%as well as the measurement precision, 
is better than the previous two continuous distributions, $G_1$ and $G_2$. Compare the three curves of Fig.~\ref{modulated_gauss}.
%of the Supplementary material.} %illustrates that there is a specific range of $a$ where $\Delta \tilde{\omega}_1 < \Delta \tilde{\omega}_0$ where this disorder model shows advantage over the disorder free situation.
To ensure that this advantage is a disorder-induced enhancement in precision of measurement,
%not merely a fluctuation of the mean value of $\Delta\tilde{\omega}_{\mathcal{Q}_{G_3}^{(p)}(t)}$, 
we have plotted the relative advantage, $\beta_{G_3}=\frac{\Delta\tilde{\omega}_0^{(p)}(t)-\Delta\tilde{\omega}^{(p)}_{\mathcal{Q}_{G_3}}}{\Delta\tilde{\omega}_0^{(p)}}$, and the standard deviation of $\Delta\tilde{\omega}^{(p)}_{\epsilon_{G_3}(t)}(t)/\Delta\tilde{\omega}_0^{(p)}(t)$, given by $\tilde{\sigma}_{G_3}$, in Fig.~\ref{modulated_gauss}-(c). It can be seen that $\beta_{G_3}$ is positive, and $\tilde{\sigma}_{G_3}$ is substantially lower than $\beta_{G_3}$.
%, and the difference between the two curves ultimately saturates to an order of $1$ with increasing value of $a$. 
So here also, we attain the phenomenon of order from disorder.

\begin{figure}
\includegraphics[width=7.5cm]{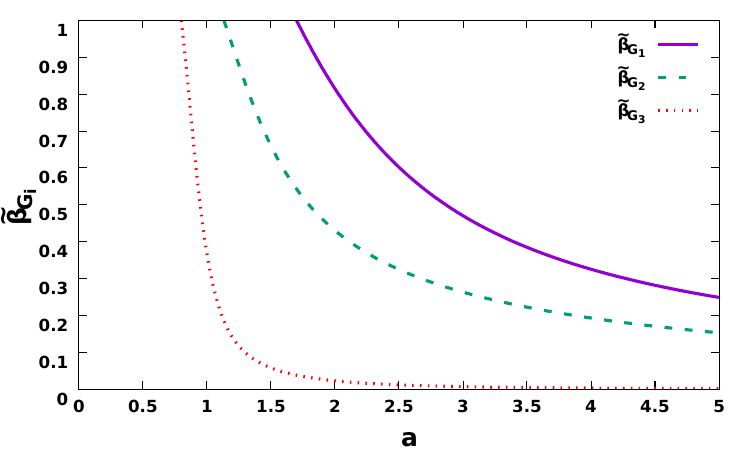}
\caption{Disorder-induced enhancement in frequency estimation for uncorrelated initial probes. Here we present the effect of modulated Gaussian glassy disorders on the minimum deviation of estimation of $\omega$, encoded in each of the initial states by evolving them with the Hamiltonian $H_1$. We depict here  the quantities $\tilde{\beta}_{G_i}=\frac{\Delta \tilde{\omega}^{(p)}_{\mathcal{Q}_{G_i}}(t)}{\Delta\tilde{\omega}_0^{(p)}(t)}$ for $i=1,2,3$, corresponding to the three modulated Gaussian distributions $G_1$, $G_2$ and $G_3$ respectively.
All  quantities plotted along both the axes are dimensionless.}
\label{projection-noi}
\end{figure}

We compare the disorder-induced advantages in frequency estimation for the distributions, $G_1$, $G_2$ and $G_3$ as  functions of the parameter $a$ in Fig.~\ref{projection-noi}. We find that the minimum error is obtained by $G_3$, followed by $G_2$ and $G_1$.

Note that not all disorder distributions  lead to benefits
in measurement precision. A discussion on the response to such disorder models in frequency estimation is provided in Appendix~\ref{appen1}. Also, an ideal system sometimes is not effected by some disorder models. The Hamiltonian $H_2$ is an example of this case. 
%See the SM for  discussions in this regard. 
We now study whether there is an enhancement in the efficiency of measurement precision, in presence of glassy disordered parameters, for $\nu$ copies of initial probes, where each copy is a two-qubit maximally entangled state. The results for cases when there is no interaction between the two probe qubits  are discussed below.
%given in the SM. 
%For the same two-qubit initial probes, in presence of external fields and interaction between the qubits, the effect of disorder on the precision of parameter estimation is discussed below.
%when the disorder parameters are chosen randomly from those distributions. 
%A few such distribution functions and the corresponding disorder-averaged minimum error in estimation of a parameter are discussed \textcolor{red}{in the Supplementary material.}
%then it is observed that there is no benefit over the disorder free scenario and $\Delta \tilde{\omega}_1/\Delta \tilde{\omega}_0>1$ always. 

\section{Measurement precision in presence of glassy disorder for copies of two-qubit maximally entangled initial probes}
 \label{Sec:6}
We now study whether there is an enhancement in the efficiency of measurement precision, in presence of glassy disordered parameters, for $\nu$ copies of initial probes, of which, each copy is a two-qubit maximally entangled state.
%We extend the single qubit case in presence of quenched disorder, to two qubits, in order to find whether the enhancement of efficiency of measurement precision, in presence of quenched disorder parameters, retain in two-qubit situations also. 
Moreover, we study the effect on the precision of incorporating interaction between the system particles and external fields in a disordered system. %consider disorder introduced in the system in presence two qubit interactions and a parallel field.
 
%\subsection{Extension of the Single Qubit case to Two Qubits:}
We consider a two-qubit system acted upon by the Hamiltonian,
\begin{equation}
    H_3= -\hbar\omega\big[(1+\epsilon_1)\ketbra{1}{1}\otimes \mathcal{I} + \mathcal{I} \otimes (1+\epsilon_2)\ketbra{1}{1}\big],
\end{equation}
where $\epsilon_1$ and $\epsilon_2$ are disorder parameters introduced in the ideal system.
The optimal input state can be constructed using Eq.~(\ref{best_input}) as $\ket{\phi_0}=(\ket{00}+\ket{11})/\sqrt{2}$ and the encoded final state after evolving with the unitary $U_3=e^{-i H_3 t}$, comes out to be $\ket{\phi_f}=(\ket{00}+e^{-i(2+\epsilon_1+\epsilon_2)\omega t}\ket{11})/\sqrt{2}$. The quantum Fisher information for this case has the form, $F_{Q}^{\epsilon_1,\epsilon_2}=t^2(2+\epsilon_1+\epsilon_2)^2$, obtained from~(\ref{pure_fq}). The single-copy scenario of the two-qubit maximally entangled initial state can be extended to $\nu$ copies of the same initial probes and  %for $\nu$ copies of the initial state. 
the error in frequency for that case turns out to be
 \begin{equation}
    \Delta \tilde{\omega}_{\mathcal{Q}}^{(e)}=\int_{\text{lb}_1}^{\text{ub}_1}\int_{\text{lb}_2}^{\text{ub}_2}\frac{1}{t\sqrt{\nu}|2+\epsilon_1+\epsilon_2|} \; P_1(\epsilon_1) P_2(\epsilon_2) d\epsilon_2 d\epsilon_1,
\end{equation}
where $\text{lb}_i$ and $\text{ub}_i$, for $i=1,2$, refer to the lower and upper bounds of the probability distribution functions $P_i(\epsilon_i)$ respectively. Here $P_1(\epsilon_1)$ and $P_2(\epsilon_2)$ are independent of each other. In absence of disorder, the deviation in frequency estimation is $\Delta\tilde{\omega}_0^{(e)}=\frac{1}{2t\sqrt{\nu}}=\frac{1}{2}\Delta\tilde{\omega}_0^{(p)}$. Therefore to attain an advantage over the ideal case, we need $\frac{\Delta\tilde{\omega}_{\mathcal{Q}}^{(e)}}{\Delta\tilde{\omega}_0^{(p)}} < \frac{1}{2}$.

Following the line pursued in case of uncorrelated initial probes, we now investigate the effect of discrete disordered systems by choosing $\epsilon_1$ and $\epsilon_2$ randomly 
%in order to determine whether the two qubit case provides an added benefit in the measurement precision. Let us first choose the disorder parameters $\epsilon_1$ and $\epsilon_2$
from the discrete disorder of type $D_1$. The corresponding probability distribution functions $P_i(\epsilon_i)$  are the same as in Eq.~(\ref{eq:D1}), with the parameters $\alpha_i$ and probability $p_i$ for $i=1,2$ respectively.
\begin{figure}
\includegraphics[width=8.5cm]{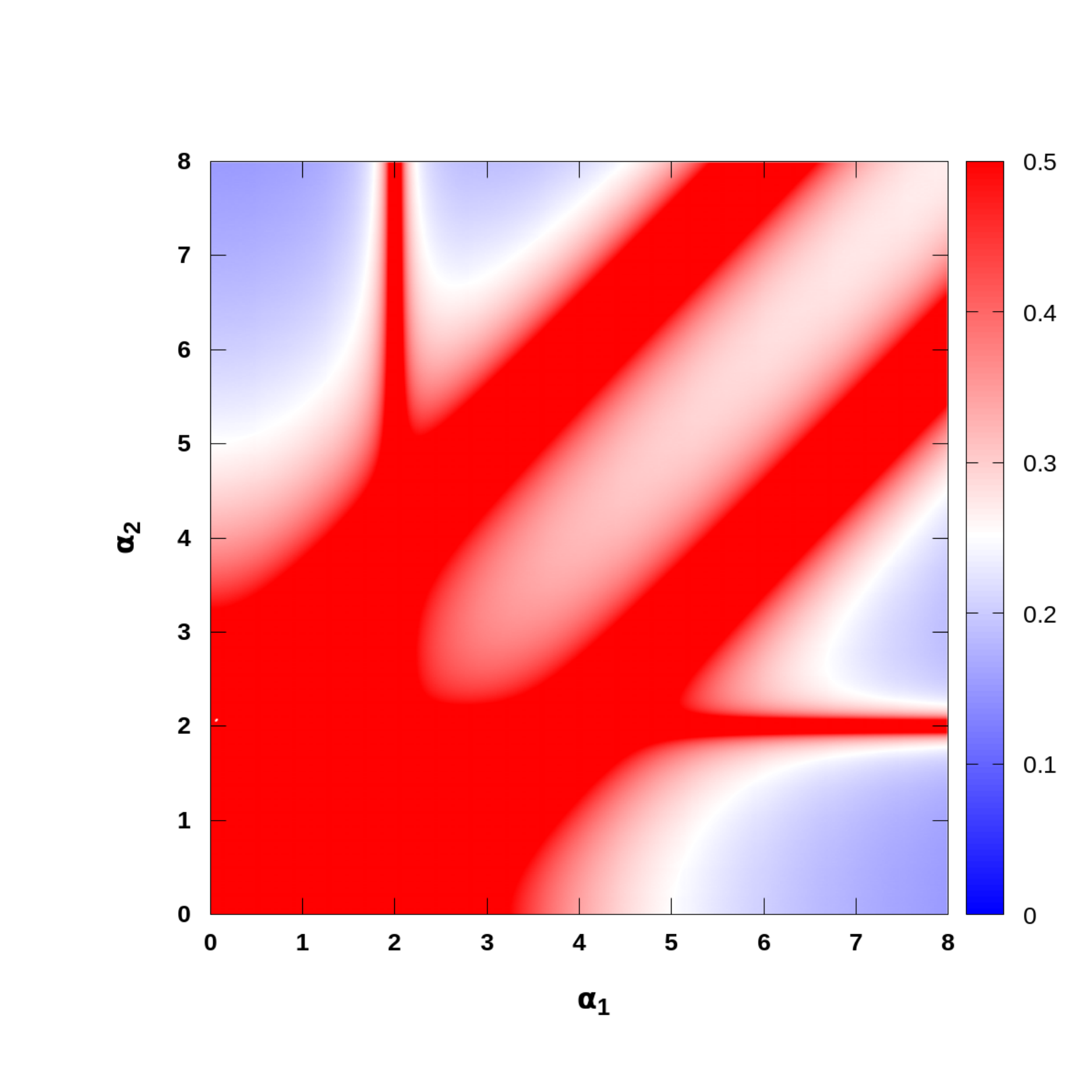}
\caption{Disorder-induced enhancement in frequency estimation for copies of two-qubit maximally entangled initial probes.  Here we depict the nature of the projection of the quantity $\frac{\Delta\tilde{\omega}_{\mathcal{Q}_{D_1}}^{(e)}}{\Delta\tilde{\omega}_0^{(p)}}$, with respect to $\alpha_1$ and $\alpha_2$, for each single copy evolving with the Hamiltonian $H_3$. Here we take $p_1=p_2=0.05$. 
%It is observed that the value of $\frac{\Delta\tilde{\omega}_2}{\Delta\tilde{\omega}_0} < \frac{1}{2}$ in a significant domain of $\alpha_x - \alpha_y$ plane suggesting an advantage over the disorder free situation. 
All the quantities plotted along both the axes are dimensionless.}
\label{projection}
\end{figure}
%Like the single qubit case, we select the disorder parameters, $\epsilon_1$ and $\epsilon_2$, from a discrete distribution of type 1, but in two dimensions, which is given by $P_x(\epsilon)P_y(\epsilon)$, where
%\begin{eqnarray}
%    P_x(\epsilon) &=& -\alpha_x \; \; \;  \text{with} \; \; \text{probability} \; \; \; \frac{(1-p_x)}{2} \nonumber \\
%                &=& 0 \; \; \; \; \; \; \; \text{with} \; \; \text{probability} \; \; \; p_x \nonumber \\
%                &=& \alpha_x \; \; \; \; \; \;  \text{with} \; \; \text{probability} \; \; \; \frac{(1-p_x)}{2}, 
%    \end{eqnarray}
%    and
%    \begin{eqnarray}
%     P_y(\epsilon) &=& -\alpha_y \; \; \;  \text{with} \; \; \text{probability} \; \; \; \frac{(1-p_y)}{2} \nonumber \\
%                &=& 0 \; \; \; \; \; \; \; \text{with} \; \; \text{probability} \; \; \; p_y \nonumber \\
%                &=& \alpha_y \; \; \; \; \; \;  \text{with} \; \; \text{probability} \; \; \; \frac{(1-p_y)}{2}, 
%    \end{eqnarray}
The minimum deviation in estimation of $\omega$ is given by
 \begin{eqnarray}
\Delta\tilde{\omega}_{\mathcal{Q}_{D_1}}^{(e)} &=&\frac{1}{t\sqrt{\nu}}\Big[\frac{1}{2} \sum_{i,j}p_i \left(1-p_j\right) \left(\frac{1}{\left| \alpha _j+2\right| }+\frac{1}{\left| 2-\alpha _j\right| }\right)\nonumber\\
&+&\frac{1}{4}\sum_{i,j}\Big\{p_i p_j+(1-p_i)(1-p_j)\Big(\frac{1}{|\alpha_i-\alpha_j+2|}\nonumber\\
&&\phantom{check korte}+\frac{1}{|(-1)^j(\alpha_i+\alpha_j)+2|}\Big)\Big\}\Big].
\end{eqnarray}
% \begin{widetext}
   %  \begin{eqnarray}
       % \Delta\tilde{\omega}_2 &=& \frac{1}{4} \left(1-p_x\right) \left(1-p_y\right) \left(\frac{1}{\left| \alpha _x-\alpha _y+2\right| }+\frac{1}{\left| -\alpha _x+\alpha _y+2\right| }+\frac{1}{\left| \alpha _x+\alpha _y+2\right| }+\frac{1}{\left| -\alpha _x-\alpha _y+2\right| }\right)\nonumber \\
      %     && +\frac{1}{2} p_x \left(1-p_y\right) \left(\frac{1}{\left| \alpha _y+2\right| }+\frac{1}{\left| 2-\alpha _y\right| }\right)+\frac{1}{2} \left(1-p_x\right) p_y \left(\frac{1}{\left| \alpha _x+2\right| }+\frac{1}{\left| 2-\alpha _x\right| }\right)+\frac{p_x p_y}{2}.
     % \end{eqnarray}
    % \end{widetext}
%of $\Delta\tilde{\omega}_2$ for this distribution 
Here $i$ and $j$ both run from $1$ to $2$. In Fig.~\ref{projection}, we plot the projection of the dimensionless quantity $\frac{\Delta\tilde{\omega}_{\mathcal{Q}_{D_1}}^{(e)}}{\Delta\tilde{\omega}_0^{(p)}}$, with respect to $\alpha_1$ and $\alpha_2$, and observe that there is a certain range of $\alpha_1,\alpha_2$ for a fixed $p_1$ and $p_2$, in which there is an advantage in the measurement precision over the ideal scenario. The advantage is highest in two regions, viz. for small $\alpha_1$ and high $\alpha_2$, and for small $\alpha_2$ and high $\alpha_1$ (see the light blue parts of Fig.~\ref{projection}). 
%\textcolor{red}{The advantage is larger for  a higher value of $\alpha_1$, $\alpha_2$ when $p_1$, $p_2$ is small. This situation is physically consistent.}
%As depicted in Fig.(\ref{2q_discrete_1}), for the particular case of $p_x=p_y=0.05$, there is a certain region in $\alpha_x-\alpha_y$ space where $\frac{\Delta\tilde{\omega}_2}{\Delta\tilde{\omega}_0} < \frac{1}{2}$.
We have also compared the standard deviation in $\frac{\Delta\tilde{\omega}_{{{(\epsilon_1)_{D_1} (\epsilon_2)_{D_1}}}}^{(e)}}{\Delta\tilde{\omega}_0^{(p)}}$, with the relative advantage, $\frac{\Delta\tilde{\omega}_0^{(p)}-\Delta\tilde{\omega}_{\mathcal{Q}_{D_1}}^{(e)}}{\Delta\tilde{\omega}_0^{(p)}}$, and notice that there exists a region in the $(\alpha_1,\alpha_2)$-plane for which the standard deviation is less than the relative advantage, which tells that in this case also, we obtain a disorder-induced enhancement in the efficiency of measurement over the ideal one. %Fig.(~\ref{projection}) depicts the projection of the quantity $\Delta\tilde{\omega}_2/\Delta\tilde{\omega}_0$ in the $\alpha_x - \alpha_y$ plane, for fixed values of $p_x$ and $p_y$, in the region where there is an advantage in presence of disorder, i.e., $\Delta\tilde{\omega}_2/\Delta\tilde{\omega}_0 < \frac{1}{2}$.
%These results are effectively an extension of the single qubit case. 
%\textcolor{red}{projection plot of $\Delta\tilde{\omega}_2$ vs $\alpha_x$, $\alpha_y$ for fixed $p_x$, $p_y$?}
Similar to the single-qubit case, the truncated Cauchy-Lorentz $(TGL)$ and truncated Gaussian $(TG)$ distributions never give an advantage over the disorder-free scenario.
%%%%%%%%%%%%%%%%%%%%%%%%%%%%%%

\section{Presence of Ising interaction between the probe qubits and introduction of external fields}
\label{Sec:7}
We now apply an external field on each of the qubits of a two-qubit single copy of the initial probe, in which the two qubits are also interacting through an Ising interaction. The disorder parameters are introduced in the coupling strengths of the interaction and the fields. %\textcolor{red}{and we estimate the interaction strength.} 
The Hamiltonian of the composite system takes the form 
\begin{eqnarray}
    H_4 &=& \overline{H}(\hat{A},\epsilon) + H_I, \;\;\;\; \text{where} \nonumber  \\
    H_I &=& \hbar Jk\big[ (1+\epsilon_1)\sigma_z^1 + (1+\epsilon_2)\sigma_z^2 \big] + \hbar J \epsilon_3 (\sigma_z^1\otimes\sigma_z^2),  \nonumber
\end{eqnarray}
with $k$ being a dimensionless constant and $\epsilon_1$, $\epsilon_2$ and $\epsilon_3$ are the disorder parameters.
%and, $k_1$ and $k_2$ are dimensionless constants corresponding to the field and the interaction strengths respectively. 
$\overline{H}(\hat{A},\epsilon)$ is an arbitrary hermitian operator which may be dependent on some other operators like $\hat{A}$, or even some disorder parameter $\epsilon$, but is independent of $J$, and also $[\overline{H}(\hat{A},\epsilon),H_I] = 0$.
%Suppose, we want to estimate the coupling constant $J$.
The parameter $J$ is encoded into the system by evolving each copy of the the initial states by the Hamiltonian $H_4$ and the best input state is evaluated using  
%The Hamiltonian undergoes a unitary evolution given by $\ket{\psi_J}=e^{-iH_4t/\hbar}\ket{\psi'_0}$, where $\ket{\psi'_0}$ is calculated using 
Eq.~\eqref{best_input}. Then the Fisher information-based lower bound, $\Delta \tilde{J}_{\mathcal{Q}}^{(2)}$, in estimation of the parameter $J$, in presence of various types of glassy disorders in the system, is evaluated numerically. Here ``$(2)$" in the superscript is to remind us that the copies of two-qubit initial probes, which can be entangled or product, are used for encoding. An important point to be noted here is that while measuring the parameter $J$ of a similar Hamiltonian containing only the interaction term, the best input state becomes a product state, instead of an maximally entangled one~\cite{Bhattacharyya}. In this situation also, the entanglement content of the initial probes depends on the relative values of disorder parameters. We will discuss about this matter after a while. 
%After that the quantum Fisher information $F_Q(J)$ is evaluated using Eq.~(\ref{pure_fq}) and the variance of $J$, $\Delta J$ follows from the Cram\'er-Rao bound. 

For the interacting two-qubit probes, governed by $H_4$, we study the effects of Gaussian and uniform disorder on the measurement precision of $J$. 
%We choose the disorder parameters randomly from the Gaussian distribution $G$ and from the uniform distribution $U$ in this section.
%respectively in Figs.~\ref{2q_gauss}-(a) and~\ref{2q_gauss}-(b).
%We 
%evaluate the error in $J$, $\Delta J_2$, and its difference from the ordered case, $(\Delta J_0-\Delta J_2)$, and
We evaluate the relative advantage in estimated $J$, viz. $\beta^{\prime}_{X}=\frac{\Delta \tilde{J}_0^{(2)}-\Delta \tilde{J}_{\mathcal{Q}_X}^{(2)}}{\Delta \tilde{J}_0^{(2)}}$, for $X$ being $G$ or $U$ (for Gaussian and uniform disorders respectively), and compare $\beta^{\prime}_{X}$ with $\sigma^{\prime}_X$, which is the standard deviation in $\frac{\Delta \tilde{J}_{(\epsilon_1)_X(\epsilon_2)_X(\epsilon_3)_X}^{(2)}}{\Delta \tilde{J}_0^{(2)}}$. Here $\Delta \tilde{J}_0^{(2)}$ is the error in $J$, estimated in absence of any type of disorder in the interaction Hamiltonian $H_I$, i.e., for $\epsilon_1=\epsilon_2=\epsilon_3=0$.
%as a function of time. 
The standard deviation, $\sigma_X^{\prime}$, for both $X=G$ and $U$, are significantly less than the corresponding $\beta^{\prime}_{X}$.
Precisely, $\sigma_G^{\prime}$ oscillates around an average value $\approx 0.33$, and the oscillation of $\beta^{\prime}_{G}$ is about $\approx 0.53$. For uniform distribution, $\sigma_U^{\prime}$ fluctuates around $\approx 0.27$, whereas the fluctuation of $\beta^{\prime}_{U}$ is around $\approx 0.44$.
In these numerical investigations the strength of the Gaussian disorder, $\sigma_G$,  is taken to be $\frac{1}{2}$ and for uniform distribution, $\sigma_U=\frac{1}{2\sqrt{3}}$, which corresponds to $s=1$. We also set $J=1.5$ and $k=0.08$.
So there are instances of order from disorder for both Gaussian and uniform glassy disorders.

Interestingly, the presence of various glassy disorders in the system provide advantages over the ideal scenario, also with respect to the entanglement content of the initial states. 
%in the optimal choice of input states in presence of disorder. Since we consider two qubit interactions, the quantum correlations between the two qubits shows an intriguing feature here. 
It can be easily checked using the optimal input state, described around Eq.~\eqref{best_input}, that for the estimation of $J$, in absence of disorder in the fields and interaction term ($\epsilon_1=\epsilon_2=\epsilon_3=0$), the best choice of initial probes are maximally entangled states, viz. $\ket{\phi_0}$. But, in presence of disorder, depending upon the relative values of $\epsilon_1$, $\epsilon_2$ and $\epsilon_3$, the optimal choice of the initial states may become a product one. It is observed that, the best initial inputs become copies of $\ket{0+}$ and $\ket{+0}$, when $\epsilon_3 < 1+\epsilon_1$ and $\epsilon_3 < 1+\epsilon_2$, respectively. 
Here $\ket{0}$ and $\ket{1}$ are eigenstates of $\sigma_z$, and $\ket{+}=(\ket{0}+\ket{1})/\sqrt{2}$.
%the best choice of input is a product state, otherwise it is maximally entangled.
So, the optimal choice of input state may switch from a maximally entangled state to a product one, in presence of disorder. The explicit form of the Hamiltonian $H_I$ shows that in absence of disorder, the Ising interaction between the two qubits disappear. This interaction comes into play in presence of disorder, which can be considered to be responsible for this peculiarity in the entanglement content of the two-qubit inputs. 
%Therefore the quantum correlations between the two qubits of the optimum input state is lower in the disordered case than in the ideal scenario. 
%Figs.~\ref{2q_gauss}-(a) and (b) depict this situation when, 
Thus, we find a situation when, even with a lower entanglement content of the initial probes, a certain metrological advantage may be obtained in glassy disordered systems over the ideal one. %It is noteworthy that there is some  added benefit in the estimate of $J$ in the presence of disorder even  with less requirement of entanglement. 
We numerically investigate the average entanglement of the optimal input states and see that it is much less the same of the ideal scenario. Since we take into account two-qubit pure states, the measure of entanglement considered is the von Neumann entropy of the local subsystem~\cite{Bennett_Popescu}. Precisely, the average entanglement for the Gaussian disorder is $\approx 0.09$, and the same for uniform disorder is $\approx 0.13$.
%The red solid curve in Fig.~\ref{entangle} illustrates the entanglement content of the optimal input state in the ideal situation, while the black and blue dashed curves represent the same when   $\epsilon_1$, $\epsilon_2$ and $\epsilon_3$ are selected at random from Gaussian and uniform distribution respectively. 

%\vspace{0.2cm}

%\noindent\emph{\textbf{Conclusion.}}-
\section{Conclusion}
\label{Sec:8}
The response to disorder in a system is typically expected to be detrimental. Here we found instances in which the effect of glassy disorders is  otherwise on the precision in quantum estimation of system parameters. 
%\textcolor{blue}{\sout{In this paper, we examined the effect of glassy disorders present in the system, on the precision in estimation of system parameters.}} 
This has potentially crucial implications for quantum devices having portions involving precision measurements, like measurement of time in atomic clocks. 

We showed  how an unbiased estimator can be identified in a disordered situation. We also found how the precision thereof can be bounded by the quantum Cr{\'a}mer-Rao inequality.

We considered copies of uncorrelated as well as two-qubit maximally entangled initial probes and identified instances where relative advantages in the efficiency of estimations are obtained in presence of glassy disorder in the system over the ideal cases. 
In looking for disorder-induced enhancement, we compared the disorder-averaged minimum Fisher information-based error with the corresponding quantity in the ideal, i.e., disorder-free, limit.  We also examined whether the difference between the disorder-averaged and ideal cases is higher than the disorder-induced fluctuations in the disorder-average.
%We find that there are some disorder models, for which, a disorder-induced enhancement in the estimation of system parameters, over the disorder-less scenarios, can be attained. These scenarios refer to the achievement of relative advantages of precision in measurement over the ideal case, which is significantly larger than the system's inherent fluctuations that occur in presence of disorders. These fluctuations can be quantified as the standard deviation of the estimated measurement error in presence of disorder. 
Moreover, we  identified cases where a glassy disorder, present in the system, reduces the requirement of initial quantum correlation - entanglement - for attaining better precision in the estimation of a parameter over an ideal - disorder free - situation, providing us a ``double advantage'', viz. order-from-disorder in conjunction with resource reduction. %Interestingly, the quantum correlations between the two qubit, measured in terms of the system's local von Neumann entropy, becomes effective in presence of interactions and the parallel field. It is noted that the correlations between the two qubits decrease when disorder is present in the system, and  one also obtains a metrological advantage in such a scenario. As a result, in the presence of disorder, there is a substantial metrological benefit when employing lesser requirement of two-qubit correlations as a resource.

%\vspace{.2cm}

%\noindent
%\emph{\textbf{Acknowledgements.}}- 

\acknowledgements
We acknowledge computations performed using Armadillo~\cite{Sanderson,Sanderson1},
%- a high quality linear algebra library for C++ language, 
%NLOPT~\cite{NLOPT}
%- a high quality algorithm (we have used 
 %(Direct-L~\cite{Gablonsky}, 
 %ISRES~\cite{Runarsson})
 %for global and 
  %Cobyla~\cite{Powell}),
  %for local optimization) for optimization procedure and 
   and QIClib~\cite{QIClib}
   %- a high quality C++ library of quantum information and computation and 
   on the cluster computing facility of Harish-Chandra Research Institute, India. This research was supported in part by the `INFOSYS scholarship for senior students'. We also acknowledge partial support from the Department of Science and Technology, Government of India through the QuEST grant (grant number DST/ICPS/QUST/Theme-3/2019/120).

\appendix
\section{Models of disorder that offer no betterment in measurement precision over the ideal scenario}
\label{appen1}
%There are some distribution functions, which shows non-beneficial results 
The disorder-averaged quantities for the Gaussian ($G$) and Cauchy-Lorentz ($CL$) distributions, represented as $\Delta \tilde{\omega}_{\mathcal{Q}_G}^{(p)}$ and $\Delta \tilde{\omega}_{\mathcal{Q}_{CL}}^{(p)}$ respectively, do not converge.
%when we choose the disorder parameter $\epsilon$ randomly from the probability distributions,  Gaussian ($G$) or Cauchy-Lorentz ($CL$) respectively. 
%The probability distribution function of Cauchy-Lorentz distribution is given in Appendix~\ref{appen1}. 
Due to this difficulty, we try truncated Gaussian ($TG$) and truncated Cauchy-Lorentz ($TCL$) distributions. A truncated distribution is one, in which the domain of definition of the random variable is reduced to a finite range.
%\textcolor{cyan}{\sout{If the disorder parameter $\epsilon$ is chosen randomly from a truncated Gaussian or truncated Cauchy-Lorentz distribution, no benefit over the ideal scenario is achieved i.e., $\Delta \tilde{\omega}^{(p)}_{\mathcal{Q}_{TG}}(t)/\Delta \tilde{\omega}_0^{(p)}(t)>1$ and $\Delta \tilde{\omega}^{(p)}_{\mathcal{Q}_{TCL}}(t)/\Delta \tilde{\omega}_0^{(p)}(t)>1$. For uniform quenched disorder ($U$) also, no disorder-induced enhancement is achieved.}} 
These distribution functions and the corresponding disorder-averaged minimum error in estimation of a parameter are discussed below. \\
%\\
%\textbf{\textbullet $\mathbf{\;\;CL}$: \textit{Cauchy-Lorentz disorder.}} This disorder is quite different from the others discussed previously. The mean does not exist for this distribution. The disorder parameter $\epsilon$ is distributed as
%\begin{equation}
%\label{eq:A1}
 %    P(\epsilon) = \frac{1}{\gamma \pi }\Big(\frac{\gamma^2}{\epsilon^2+\gamma^2} \Big), \;\;\; \text{where}  \;\;\; -\infty < \epsilon \le \infty .
%\end{equation}
%For uncorrelated initial probes, the disorder-averaged error in frequency estimation diverges for this distribution. \\
\\
\textbf{\textbullet $\mathbf{\;\;TCL}$: \textit{Truncated Cauchy-Lorentz disorder.}}
If we truncate the range of the parameter $\epsilon$ within a finite range in~(\ref{eq:A1}), we get a truncated Cauchy-Lorentz distribution, given by
\begin{eqnarray}
    P(\epsilon) &=& \frac{1}{\gamma \pi A_{TCL} }\Big(\frac{\gamma^2}{\epsilon^2+\gamma^2} \Big), \;\;\; \text{where}  \;\;\; -a < \epsilon \le a, \nonumber \\
    &=& 0 \quad \quad \quad \quad \quad \quad \quad \quad \; \text{otherwise}.
\end{eqnarray}
The normalisation constant $A_{TCL} = \frac{2}{\gamma}\tan^{-1}\frac{a}{\gamma}$.
The disorder-averaged error in frequency estimation for $\nu$ copies of uncorrelated initial probes for this disorder is
\begin{equation}
\label{Cauchy_Lorentz}
\Delta\tilde{\omega}^{(p)}_{\mathcal{Q}_{TCL}} = \frac{1}{t\sqrt{\nu}}\frac{\gamma}{1+\gamma^2} \Big(\frac{\ln\big|\frac{1+a}{1-a} \big|}{2\tan^{-1}\frac{a}{\gamma}} + \frac{1}{\gamma}\Big).
\end{equation}
This equation always gives $\Delta \tilde{\omega}^{(p)}_{\mathcal{Q}_{TCL}}(t)/\Delta \tilde{\omega}_0^{(p)}(t)>1$ for $\gamma > 0$ and $0<a<1$.\\
\\
\textbf{\textbullet $\mathbf{\;\;TG}$: \textit{Truncated Gaussian disorder.}}
In a similar way, 
%this distribution is also obtained by truncating the range of the disorder parameter of the usual Gaussian distribution denoted by $G$. So, the truncated Gaussian distribution is 
a truncated Gaussian distribution can be defined as
 \begin{eqnarray}
P(\epsilon) &=& \frac{1}{A_{TG}\sigma_{TG}\sqrt{2\pi}} e^{-(\epsilon/\sigma_{TG}\sqrt{2})^2}, \;\; \text{when} \;\; -a \le \epsilon \le a, \nonumber \\
&=& 0\quad \quad \quad \quad \quad \quad \quad \quad \quad \quad \quad\quad\,\text{otherwise}.
    \end{eqnarray}
Here $A_{TG}$ is the normalization constant given by $A_{TG}=\text{erf}(\frac{a}{\sqrt{2}\sigma_{TG}})$. The disorder-averaged deviation in frequency in presence of this disorder is evaluated numerically for $\nu$ copies of uncorrelated initial probes and the result does not provide advantage in the precision of measurement over the ideal situation.\\
\\
%It has been numerically observed that this distribution also gives an error in frequency greater than unity in the single qubit case.
\textbf{\textbullet $\mathbf{\;\;U}$: \textit{Uniform disorder.}}  
%Although the uniform disorder $U$, discussed in \textcolor{red}{the main text}, provides an advantage in the efficiency of measurement of a system parameter over the disorder-free situation, In case of two-qubit copies taken as initial probes in presence of field and interactions, 
The uniform disorder mentioned in Eq.~(\ref{uniform}) does not provide an advantage in the efficiency of measurement of a system parameter over the disorder-free situation when uncorrelated initial probes are taken.
%It reduces the measurement precision when uncorrelated initial probes are taken. 
The disorder-averaged error in estimation of frequency for $\nu$ copies of uncorrelated initial states comes out to be
\begin{equation}
    \Delta\tilde{\omega}^{(p)}_{\mathcal{Q}_U} =\frac{1}{t\sqrt{\nu}} \frac{1}{s} \ln\big|\frac{2+s}{2-s} \big|,
\end{equation}
which is always $> \Delta \tilde{\omega}_0^{(p)}$.\\
\\An important point to be noted is that while an introduction of disorder parameter in the scale of energy, as in $H_1$, leads to a definite change in the frequency estimation, there is no change in the same if we introduce the same disorder parameter in the form of an energy shift (see $H_2$). The reason is the following. For a unitary evolution with the unitary $U_2=e^{-iH_2 t/\hbar}$,
%, which can be written as $e^{i\tilde{\omega} \ketbra{1}{1}t}e^{i\epsilon \ketbra{1}{1} t}$ as $\big[\tilde{\omega} \ketbra{1}{1},\epsilon \ketbra{1}{1}\big]=0$.
the variance in estimating $\omega$ is given by $\frac{1}{4\nu\Delta^2\mathcal{H}_1}$,
%\begin{equation}
%\label{H0_crbound}
 %   \Delta^2\tilde{\omega}^{(p)} \ge \frac{1}{4\nu\Delta^2\mathcal{H}_1},
%\end{equation}
where $\mathcal{H}_1$ is the ideal-system Hamiltonian. Hence, in this case, the error is independent of $\epsilon$, and so there is no effect, on metrological precision of frequency estimation, of having this type of disorder in the system. 

%\section{Comparison of the disorder-induced advantage for distributions $G_1$,$G_2$, $G_3$}
%%%%%%%%%%%%%%%%%%%%%%%%%%%%%%%%%%%%%%%%%%%%%%%%%%%%%%%%%%%%%%%%%%%%%%%%%%%%%%%%%%%%


\begin{thebibliography}{100}
\bibitem{Braunstein1} S. L. Braunstein, \textit{Quantum limits on precision measurements of phase},
Phys. Rev. Lett. \textbf{69}, 3598 (1992). 

\bibitem{Lloyd0} V. Giovannetti, S. Lloyd and L. Maccone, \textit{Quantum Metrology},
Phys. Rev. Lett. \textbf{96}, 010401 (2006).
\bibitem{Treutlein} L. Pezz\`{e}, A. Smerzi, M. K. Oberthaler, R. Schmied and
P. Treutlein, \textit{Quantum metrology with nonclassical states
of atomic ensembles}, Rev. Mod. Phys. \textbf{90}, 035005 (2018).

%\bibitem{Cirac&Plenio} S.F Huelga, C. Macchiavello, T. Pellizzari, A.K. Ekert, M. B. Plenio and J.I. Cirac, \textit{The Improvement of Frequency Stardards with Quantum Entanglement}, Phys. Rev. Lett. \textbf{79}, 3865 (1997).

\bibitem{q_imaging} I. Ruo-Berchera, A. Meda, E. Losero, A. Avella,
N. Samantaray and M. Genovese, \textit{Improving resolution sensitivity trade off in sub-shot noise quantum imaging}, Applied Phys Lett. \textbf{116}, 214001 (2020).
\bibitem{Genoni3} F. Albarelli, M. Barbieri, M. G. Genoni and I. Gianani, \textit{A perspective on multiparameter quantum metrology: from theoretical tools to applications in quantum imaging}, Phys. Lett. A \textbf{384}, 126311 (2020).
\bibitem{Liang}  K. Liang, S. A. Wadood and A. N. Vamivakas, \textit{Coherence effects on estimating two-point separation}, Optica \textbf{8}, 243 (2021).

\bibitem{therm1} M. T. Mitchison, T. Fogarty, G. Guarnieri, S. Campbell,
T. Busch and J. Goold, \textit{In situ thermometry of a cold
fermi gas via dephasing impurities}, Phys. Rev. Lett. \textbf{125},
080402 (2020).
\bibitem{therm2} M. R. J\o{}rgensen, P. P. Potts, M. G. A. Paris and J. B.
Brask, \textit{Tight bound on finite-resolution quantum thermometry at low temperatures}, Phys. Rev. Research \textbf{2}, 033394 (2020).

\bibitem{gravity} F. Acernese \textit{et al.} (Virgo Collaboration), \textit{Increasing the
astrophysical reach of the advanced virgo detector via
the application of squeezed vacuum states of light}, Phys.
Rev. Lett. \textbf{123}, 231108 (2019).

\bibitem{mag1} J. F. Barry, J. M. Schloss, E. Bauch, M. J. Turner, C. A.
Hart, L. M. Pham and R. L. Walsworth, \textit{Sensitivity optimization for nv-diamond magnetometry}, Rev. Mod. Phys. \textbf{92}, 015004 (2020).
\bibitem{mag2} R. Patel, L. Zhou, A. Frangeskou, G. Stimpson,
B. Breeze, A. Nikitin, M. Dale, E. Nichols, W. Thornley, B. Green, M. Newton, A. Edmonds, M. Markham,
D. Twitchen and G. Morley, \textit{Subnanotesla magnetometry with a fiber-coupled diamond sensor}, Phys. Rev. Applied \textbf{14}, 044058 (2020).

 \bibitem{Holland} M. J.Holland and K. Burnett,  \textit{Interefometric Detection
of Optical Phase Shift at the Heisenberg Limit}, Phys.
Rev. Lett. \textbf{71}, 1355 (1993).
 \bibitem{opt7} Z. Hradil, R. My\v{s}ka, J. Pe\v{r}ina, M. Zawisky, Y. Hasegawa and H. Rauch, \textit{Quantum Phase in Interferometry},
Phys. Rev. Lett. \textbf{76}, 4295 (1996).
\bibitem{Sanders} B. C. Sanders and G. J. Milburn, \textit{Optimal Quantum Measurements for Phase Estimation}, Phys. Rev. Lett. \textbf{75}, 2944 (1995).
\bibitem{Ou} Z. Y. Ou, \textit{Complementarity and Fundamental Limit in Precision Phase Measurement}, Phys. Rev. Lett. \textbf{77}, 2352 (1996).
 \bibitem{Dowling} H. Lee, P. Kok and J. P. Dowling, \textit{A Quantum Rosetta Stone for Interferometry},
J. Mod. Opt. \textbf{49}, 2325 (2002).

\bibitem{Heinzen} D. J. Wineland, J. J. Bollinger, W. M. Itano, F. L. Moore and D. J. Heinzen, \textit{Spin squeezing and reduced quantum noise in spectroscopy}, Phys. Rev. A \textbf{46}, R6797(R) (1992). 

\bibitem{Cirac&Plenio} S.F Huelga, C. Macchiavello, T. Pellizzari, A.K. Ekert, M. B. Plenio and J.I. Cirac, \textit{The Improvement of Frequency Stardards with Quantum Entanglement}, Phys. Rev. Lett. \textbf{79}, 3865 (1997).

\bibitem{Maccone} V. Giovannetti, S. Lloyd and L. Maccone, \textit{Advances in Quantum Metrology}, Nature Photonics \textbf{5}, 222 (2011).
\bibitem{Giovannetti_Maccone} L. Maccone and V. Giovannetti, \textit{Beauty and the noisy
beast}, Nature Physics \textbf{7}, 376 (2011).
\bibitem{Pirandola} D. Braun, G. Adesso, F. Benatti, R. Floreanini, U. Marzolino, M. W. Mitchell and S. Pirandola, \textit{Quantum enhanced measurements without entanglement}, Rev. Mod. Phys. \textbf{90}, 035006 (2018).
\bibitem{Plenio1} M. B. Plenio and S. Virmani, \textit{An Introduction to entanglement measures},  Quant. Inf. Comput. \textbf{7}, 1 (2007). 
\bibitem{Horodecki} R. Horodecki, P. Horodecki, M. Horodecki and K. Horodecki, \textit{Quantum entanglement}, Rev. Mod. Phys. \textbf{81}, 865 (2009).
\bibitem{Toth} O. G{\"u}hne and G. T{\'o}th, \textit{Entanglement detection}, Physics Reports \textbf{474}, 1 (2009).

\bibitem{Lewenstein1} S. Das, T. Chanda, M. Lewenstein, A. Sanpera, A. Sen(De) and U. Sen, \textit{The separability versus entanglement problem}, in \emph{Quantum Information}, edited by D. Bru{\ss} and G. Leuchs (Wiley, Weinheim, 2019), chapter 8.

\bibitem{Lukin} A. Andr\'{e}, A. S. S\o{}rensen and M. D. Lukin, \textit{Stability of
Atomic Clocks Based on Entangled Atoms}, Phys. Rev.
Lett. \textbf{92}, 230801 (2004).
\bibitem{Nagata} T. Nagata, R. Okamoto, J. L. O’Brien, K. Sasaki and
S. Takeuchi, \textit{Beating the standard quantum limit with
four-entangled photons}, Science \textbf{316}, 726 (2007).


\bibitem{Roy} S. M. Roy and S. L. Braunstein, \textit{Exponentially Enhanced
Quantum Metrology}, Phys. Rev. Lett. \textbf{100}, 220501
(2008).
\bibitem{Lewenstein} R. Augusiak, J. Ko lody\'{n}ski, A. Streltsov, M. N. Bera,
A. Ac\'{i}n and M. Lewenstein, \textit{Asymptotic role of entanglement in quantum metrology}, Phys. Rev. A \textbf{94}, 012339 (2016).

\bibitem{Matsuzaki} L. B. Ho, H. Hakoshima, Y. Matsuzaki, M. Matsuzaki
and Y. Kondo, \textit{Multiparameter quantum estimation under dephasing noise}, Phys. Rev. A \textbf{102}, 022602 (2020).
\bibitem{Hu} X. Deng, S.L. Chen, M. Zhang, X.F. Xu, J. Liu, Z. Gao, X.C. Duan, M.K. Zhou, L. Cao and Z. K. Hu, \textit{Quantum metrology with precision reaching beyond-1/N scaling through N-probe entanglement generating interactions}, 	Phys. Rev. A \textbf{104}, 012607 (2021)


\bibitem{opt1} E. Yablonovich and R.B Vrijen, \textit{Optical projection lithography at half the Rayleigh resolution limit by two-photon exposure}, Opt. Eng. \textbf{38}, 334 (1999).
\bibitem{opt2} A. N. Boto, P. Kok, D. S. Abrams, S. L. Braunstein, C. P. Williams and J. P. Dowling, \textit{Quantum Interferometric Optical Lithography: Exploiting Entanglement to Beat the Diffraction Limit}, Phys. Rev. Lett. \textbf{85}, 2733 (2000).
\bibitem{opt3} P. Kok, A. N. Boto, D. S. Abrams, C. P. Williams, S. L. Braunstein and J. P. Dowling, \textit{Quantum-interferometric optical lithography:Towards arbitrary two-dimensional patterns},
 Phys. Rev. A \textbf{63}, 063407 (2001).
\bibitem{opt4} G. S. Agarwal, M. O. Scully and H. Walther, \textit{Inhibition of Decoherence due to Decay in a Continuum},
 Phys. Rev. Lett. \textbf{86}, 1389 (2001).
\bibitem{opt5} M. D'Angelo, M. V. Chekhova and  Y. Shih, \textit{Two-Photon Diffraction and Quantum Lithography},
 Phys. Rev. Lett. \textbf{87}, 013602 (2001).
\bibitem{opt6} A. Rauschenbeutel, P. Bertet, S. Osnaghi, G. Nogues, M. Brune, J. M. Raimond and S. Haroche, \textit{Controlled entanglement of two field modes in a cavity quantum electrodynamics experiment},
 Phys. Rev. A \textbf{64}, 043802 (2001).

\bibitem{Giovannetti_Maccone2} V. Giovannetti, S. Lloyd and L. Maccone, \textit{Quantum-Enhanced Measurements: Beating the Standard Quantum Limit}, 	Science \textbf{306}, 1330 (2004).

\bibitem{Cramer}  H. Cram\'{e}r, \textit{Mathematical Methods of Statistics} (Princeton University, Princeton NJ, 1946).

\bibitem{Helstrom} C. W. Helstrom, \textit{Quantum Detection and Estimation Theory}
(Academic Press, New York, 1976).
\bibitem{Holevo} A. S. Holevo,  \textit{Probabilistic and Statistical Aspect of
Quantum Theory} (North-Holland, Amsterdam, 1982).
\bibitem{Braunstein} S. L. Braunstein and C. M. Caves, \textit{Statistical distance and the geometry of quantum states
}, Phys. Rev. Lett. \textbf{72}, 3439
(1994).
\bibitem{Milburn} S. L. Braunstein, M. C. Caves and G. J. Milburn, 
\textit{Generalized Uncertainty Relations: Theory, Examples,
and Lorentz Invariance}, Ann.  Phys. \textbf{247}, 135
(1996).
\bibitem{Hayashi1} M. Hayashi, \textit{Asymptotic theory of quantum statistical inference: selected papers}  (World Scientific Publishing, 2005).
\bibitem{Hayashi2} M. Hayashi,  \textit{Quantum Information} (Springer, Berlin,
2006).
\bibitem{Paris_book}  M. G. A. Paris, \textit{Quantum estimation for quantum technology}, Int. J. Quantum Inf. \textbf{7}, 125 (2009).
\bibitem{Paris4} L. Seveso, M. A. C. Rossi and M. G. A. Paris, \textit{Quantum metrology beyond the Quantum Cram\'{e}r-Rao theorem}, Phys. Rev. A \textbf{95}, 012111 (2017).

\bibitem{review1} Kok Chuan Tan, Hyunseok Jeong, \textit{Nonclassical Light and Metrological Power: An Introductory Review}, AVS Quantum Sci. \textbf{1}, 014701 (2019).

\bibitem{Saunders} J. Saunders and J. F. V. Huele, \textit{Qubit Quantum Metrology with Limited Measurement Resources}, arXiv:2108.02876.

\bibitem{Shaji} A. Shaji and C. M. Caves,  \textit{Qubit metrology and decoherence}, Phys. Rev. A \textbf{76}, 032111 (2007).
\bibitem{Genoni} F. Albarelli, M. A. C. Rossi, D. Tamascelli and M. G. Genoni, \textit{Restoring Heisenberg scaling in noisy quantum metrology by monitoring the environment}, Quantum \textbf{2},
110 (2018).
\bibitem{Wang} Y. Che, J. Liu, X.-M. Lu and X. Wang, \textit{Multiqubit matter-wave interferometry under decoherence and the heisenberg scaling recovery}, Phys. Rev. A \textbf{99}, 033807 (2019).
\bibitem{Binder} P. Binder and D. Braun, \textit{Quantum parameter estimation of the frequency and damping of a harmonic oscillator}, Phys. Rev. A \textbf{102}, 012223 (2020).
\bibitem{Agarwal} J. Wang, L. Davidovich and G. S. Agarwal, \textit{Quantum sensing of open systems: Estimation of damping constants and temperature}, Phys. Rev. Res. \textbf{2}, 033389
(2020).
\bibitem{Huelga2} A. Smirne, J. Kolody\'{n}ski, S. F. Huelga and
R. Demkowicz-Dobrza\'{n}ski, \textit{Ultimate precision limits for
noisy frequency estimation}, Phys. Rev. Lett. \textbf{116}, 120801.
(2016).

\bibitem{Paris3} A. Monras and M. G. A. Paris, \textit{Optimal quantum estimation of loss in bosonic channels}, Phys. Rev. Lett. \textbf{98}, 160401 (2007).
\bibitem{Kurizki} A. Zwick, G. A. Alvarez, and G. Kurizki, \textit{Maximizing information on the environment by dynamically controlled qubit probes}, Phys. Rev. Applied \textbf{5}, 014007 (2016).
\bibitem{Huelga} J. F. Haase, A. Smirne, J. Kolody\'{n}ski, R. DemkowiczDobrza\'{n}ski and S. F. Huelga, \textit{Fundamental limits to frequency estimation: a comprehensive microscopic perspective}, New Journal of Physics \textbf{20}, 053009 (2018).
\bibitem{Paris}M. Bina, F. Grasselli and M. G. A. Paris, \textit{Continuousvariable quantum probes for structured environments}, Phys. Rev. A \textbf{97}, 012125 (2018).

\bibitem{Paris2}  D. Tamascelli, C. Benedetti, H.-P. Breuer and M. G. A. Paris, \textit{Quantum probing beyond pure dephasing}, New Journal of Physics \textbf{22}, 083027 (2020).



\bibitem{Aharony} A. Aharony, \textit{Spin-flop multicritical points in systems with random fields and in spin glasses}, Phys. Rev. B \textbf{18}, 3328 (1978).
\bibitem{Misguich} G. Misguich and C. Lhuillier, \textit{Two-dimensional quantum antiferromagnets, in frustrated spin systems}, edited by H. T. Diep
(World-Scientific, Singapore, 2005).
\bibitem{Villain} J. Villain, R. Bidaux, J.P. Carton, and R. Conte, \textit{Order as an
effect of disorder}, J. Physique \textbf{41}, 1263 (1980).
\bibitem{Minchau} B. J. Minchau and R. A. Pelcovits, \textit{Two-dimensional XY model in a
random uniaxial field}, Phys. Rev. B \textbf{32}, 3081 (1985).
\bibitem{Henley} C. L. Henley, \textit{Ordering due to disorder in a frustrated vector antiferromagnet}, Phys. Rev. Lett. \textbf{62}, 2056 (1989).
\bibitem{Moreo} A. Moreo, E. Dagotto, T. Jolicoeur, and J. Riera, $Cu^{63}$  \textit{Knight shifts in the superconducting state of Y} $Ba_2$ $Cu_3$ $O_{7-{\delta}}$ $(T_c=90 K)$, Phys. Rev. B \textbf{42}, 6283 (1990). 
\bibitem{Feldman} D. E. Feldman, \textit{Exact zero-temperature
critical behavior of the ferromagnet in the uniaxial random
field}, J. Phys. A \textbf{31}, L177 (1998).
\bibitem{Volovik} G. E. Volovik, \textit{Random
anisotropy disorder in superfluid 3He-A in aerogel}, JETP Lett.
\textbf{84}, 455 (2006). 
\bibitem{Abanin}D. A. Abanin, P. A. Lee and L. S. Levitov,
\textit{Randomness-induced XY ordering in a graphene quantum hall
ferromagnet}, Phys. Rev. Lett. \textbf{98}, 156801 (2007).
\bibitem{Adamska}L. Adamska, M. B. Silva Neto and C. Morais Smith, $La_{2-x} Sr_x Cu_{1-z} Zn_z O_4$ \textit{: Role of Dzyaloshinskii-Moriya and XY anisotropies}, Phys. Rev. B \textbf{75}, 134507 (2007).
\bibitem{Santos} L. F. Santos, G. Rigolin and C. O. Escobar, \textit{Entanglement
versus chaos in disordered spin chains}, Phys. Rev. A \textbf{69},
042304 (2004).
\bibitem{Benneti} C. Mej\'{i}a-Monasterio, G. Benenti, G. G. Carlo
and G. Casati, \textit{Entanglement across a transition to quantum
chaos}, Phys. Rev. A \textbf{71}, 062324 (2005). 
\bibitem{Lakshminarayan} A. Lakshminarayan
and V. Subrahmanyam, \textit{Multipartite entanglement in a onedimensional time-dependent Ising model}, Phys. Rev. A \textbf{71},
062334 (2005).
\bibitem{Karthik}J. Karthik, A. Sharma and A. Lakshminarayan, \textit{Entanglement, avoided crossings, and quantum chaos
in an Ising model with a tilted magnetic field}, Phys. Rev. A \textbf{75},
022304 (2007);.
\bibitem{Brown}W. G. Brown, L. F. Santos, D. J. Sterling and
L. Viola, \textit{Quantum chaos, delocalization, and entanglement
in disordered heisenberg models}, Phys. Rev. E \textbf{77}, 021106
(2008).
\bibitem{Dukesz}F. Dukesz, M. Zilbergerts and L. F. Santos, \textit{Interplay between interaction and (un)correlated disorder in onedimensional many-particle systems: delocalization and global
entanglement}, New J. Phys. \textbf{11}, 043026 (2009).
\bibitem{Hide}J. Hide, W. Son and V. Vedral, \textit{Enhancing the detection of natural thermal entanglement with disorder}, Phys. Rev. Lett. \textbf{102}, 100503
(2009).
\bibitem{Fujii}K. Fujii and K. Yamamoto, \textit{Anti-Zeno effect for quantum transport in disordered systems}, Phys. Rev. A \textbf{82}, 042109
(2010).
\bibitem{Prabhu} R. Prabhu, S. Pradhan, A. Sen(De) and U. Sen, \textit{Disorder overtakes order in information concentration over quantum networks}, Phys. Rev. A \textbf{84}, 042334 (2011).
\bibitem{Munoz} P. V. Mart\'{i}n, J. A. Bonachela and M. A. Munoz, \textit{Quenched 
disorder forbids discontinuous transitions in nonequilibrium
low-dimensional systems}, Phys. Rev. E \textbf{89}, 012145 (2014).
\bibitem{Sacha} A. Niederberger, T. Schulte, J. Wehr, M. Lewenstein, L.
Sanchez-Palencia and K. Sacha, \textit{Disorder-induced order in
two-component Bose-Einstein condensates}, Phys. Rev. Lett.
\textbf{100}, 030403 (2008).
\bibitem{Sacha2}A. Niederberger, J. Wehr, M. Lewenstein
and K. Sacha, \textit{Disorder-induced phase control in superfluid
Fermi-Bose mixtures}, Europhys. Letts. \textbf{86}, 26004 (2009).
\bibitem{Wehr} A. Niederberger, M. M. Rams, J. Dziarmaga, F. M. Cucchietti, J.
Wehr and M. Lewenstein, \textit{Disorder-induced order in quantum
XY chains}, Phys. Rev. A \textbf{82}, 013630 (2010).
\bibitem{Osborne} D. I. Tsomokos, T. J. Osborne and C. Castelnovo, \textit{Interplay of topological order and spin glassiness in the toric code under random magnetic fields}, Phys. Rev. B \textbf{83}, 075124 (2011).
\bibitem{Foster} M. S. Foster, H.-Y. Xie and Y.-Z. Chou, \textit{Topological protection, disorder, and interactions: survival at the surface of three-dimensional
topological superconductors}, Phys. Rev. B \textbf{89}, 155140 (2014).

\bibitem{Anderson}P. W. Anderson, \textit{Absence of diffusion in certain random lattices}, Phys. Rev. \textbf{109}, 1492 (1958).
\bibitem{Brout} R. Brout, \textit{Statistical mechanical theory of a random ferromagnetic system}, Phys. Rev. \textbf{115}, 824 (1959).
\bibitem{Thomas} J. Kurmann, H. Thomas and G. M\"{u}ller, \textit{Antiferromagnetic
long-range order in the anisotropic quantum spin chain}, Phys. A: Stat. Mech. \textbf{112}, 235 (1982).
\bibitem{Anderson1} P. W. Anderson, \textit{Basic notions of condensed matter physics},
Frontiers in Physics \textbf{55}, 49 (1984).
\bibitem{Lee} P. A. Lee and T. V. Ramakrishnan, \textit{Disordered electronic systems}, Rev. Mod. Phys. \textbf{57}, 287 (1985).
\bibitem{Young} K. Binder and A. P. Young, \textit{Spin glasses: Experimental facts,
theoretical concepts, and open questions}, Rev. Mod. Phys. \textbf{58}, 801 (1986).
%[94] D. Chowdhury, Spin Glasses and Other Frustrated Systems
%(Co-published with Princeton university press, 1986).
%[95] M. Mezard, G. Parisi, M. A. Virasoro, and D. J. Thouless, Spin
%glass theory and beyond, Physics Today 41, 109 (1988).
%[96] C. L. Henley, Ordering due to disorder in a frustrated vector
%antiferromagnet, Phys. Rev. Lett. 62, 2056 (1989).
\bibitem{Albert} A. L. Barab\'{a}si and R. Albert, \textit{Emergence of scaling in random
networks}, Science \textbf{286}, 509 (1999).
\bibitem{Goltsev} A. V. Goltsev, S. N. Dorogovtsev and J. F. F. Mendes, \textit{Critical
phenomena in networks}, Phys. Rev. E \textbf{67}, 026123 (2003).
\bibitem{Ahufinger} V. Ahufinger, L. Sanchez-Palencia, A. Kantian, A. Sanpera
and M. Lewenstein, \textit{Disordered ultracold atomic gases in optical lattices: A case study of Fermi-Bose mixtures}, Phys. Rev. A \textbf{72}, 063616 (2005).
%[100] G. E. Volovik, Random anisotropy disorder in superfluid 3HeA in aerogel, JETP Letters 84, 455 (2006).
%[101] C. De Dominicis and I. Giardina, Random fields and spin
%glasses: a field theory approach (Cambridge University Press,
%2006).
%[102] L. Adamska, M. B. Silva Neto, and C. Morais Smith,
%Competing impurities and reentrant magnetism in
%La2−xSrxCu1−zZnzO4: Role of Dzyaloshinskii-Moriya and
%XY anisotropies, Phys. Rev. B 75, 134507 (2007).
%[103] D. A. Abanin, P. A. Lee, and L. S. Levitov, Randomness induced XY ordering in a graphene quantum hall ferromagnet,
%Phys. Rev. Lett. 98, 156801 (2007).
\bibitem{Chakrabarti} A. Das and B. K. Chakrabarti, \textit{Colloquium: Quantum annealing and analog quantum computation}, Rev. Mod. Phys. \textbf{80}, 1061 (2008).
\bibitem{Fallani} L. Fallani, C. Fort and M. Inguscio, \textit{Bose–einstein condensates in disordered potentials},  Adv. At. Mol. Opt.  \textbf{56} (Academic Press, 2008) pp. 119–160.
%[106] A. Niederberger, T. Schulte, J. Wehr, M. Lewenstein,
%L. Sanchez-Palencia, and K. Sacha, Disorder-induced order
%in two-component bose-einstein condensates, Phys. Rev. Lett.
%100, 030403 (2008).
\bibitem{Inguscio} A. Aspect and M. Inguscio, \textit{Anderson localization of ultracold
atoms}, Phys. Today \textbf{62}, 30 (2009).
\bibitem{Alloul} H. Alloul, J. Bobroff, M. Gabay and P. J. Hirschfeld, \textit{Defects
in correlated metals and superconductors}, Rev. Mod. Phys. \textbf{81}, 45 (2009).
%[109] J. Hide, W. Son, and V. Vedral, Enhancing the detection of
%natural thermal entanglement with disorder, Phys. Rev. Lett.
%102, 100503 (2009).
\bibitem{Wehr_Sacha} A. Niederberger, J. Wehr, M. Lewenstein and K. Sacha,
\textit{Disorder-induced phase control in superfluid fermi-bose mixtures}, EPL \textbf{86}, 26004 (2009). 
\bibitem{Sanchez} L. Sanchez-Palencia and M. Lewenstein, \textit{Disordered quantum
gases under control}, Nat. Phys.  \textbf{6}, 87 (2010).
\bibitem{Modugno} G. Modugno, \textit{Anderson localization in bose–einstein condensates}, Rep. Prog. Phys. \textbf{73}, 102401 (2010).
%[113] A. Niederberger, M. M. Rams, J. Dziarmaga, F. M. Cucchietti, J. Wehr, and M. Lewenstein, Disorder-induced order in
%quantum XY chains, Phys. Rev. A 82, 013630 (2010).
%[114] D. I. Tsomokos, T. J. Osborne, and C. Castelnovo, Interplay
%of topological order and spin glassiness in the toric code under
%random magnetic fields, Phys. Rev. B 83, 075124 (2011).
\bibitem{Aurbach} A. Auerbach, \textit{Interacting electrons and quantum magnetism}, (Springer Science and Business Media, 2012).
\bibitem{Shapiro} B. Shapiro, \textit{Cold atoms in the presence of disorder}, J. Phys. A Math. Theor. \textbf{45}, 143001 (2012).
\bibitem{Hide_new} J. Hide, \textit{Concurrence in disordered systems},J. Phys. A Math. Theor. \textbf{45}, 115302 (2012).
\bibitem{Alvarez} J. P. \'{A}lvarez Z\'{u}$\tilde{n}$iga and N. Laflorencie, Bose-glass transition
and spin-wave localization for 2d bosons in a random potential, Phys. Rev. Lett. \textbf{111}, 160403 (2013).
%[119] M. S. Foster, H.-Y. Xie, and Y.-Z. Chou, Topological protection, disorder, and interactions: Survival at the surface of
%three-dimensional topological superconductors, Phys. Rev. B
%89, 155140 (2014).
%[120] P. Villa Martín, J. A. Bonachela, and M. A. Muñoz, Quenched
%disorder forbids discontinuous transitions in nonequilibrium
%low-dimensional systems, Phys. Rev. E 89, 012145 (2014).
\bibitem{Havlin} A. Majdandzic, B. Podobnik, S. V. Buldyrev, D. Y. Kenett,
S. Havlin and H. Eugene Stanley, \textit{Spontaneous recovery in dynamical networks}, Nat. Phys.  \textbf{10}, 34 (2014).

\bibitem{Pollmann} J. A. Kj$\ddot{a}$ll, J. H. Bardarson and F. Pollmann, \textit{Many-body
localization in a disordered quantum Ising chain}, Phys. Rev. Lett. \textbf{113}, 107204 (2014).
\bibitem{Yao} Z. Yao, K. P. C. da Costa, M. Kiselev and N. Prokof’ev, \textit{Critical exponents of the superfluid–bose-glass transition in three
dimensions}, Phys. Rev. Lett. \textbf{112}, 225301 (2014).
\bibitem{Eisert} J. Eisert, M. Friesdorf and C. Gogolin, \textit{Quantum many-body
systems out of equilibrium}, Nat. Phys. \textbf{11}, 124 (2015).
\bibitem{Fanchini} B. Cakmak, G. Karpat and F. F. Fanchini, \textit{Factorization and
criticality in the anisotropic XY chain via correlations}, Entropy \textbf{17}, 790 (2015).
\bibitem{Prabhu2} D. Sadhukhan, R. Prabhu, A. Sen(De) and U. Sen, \textit{Quantum
correlations in quenched disordered spin models: Enhanced
order from disorder by thermal fluctuations}, Phys. Rev. E \textbf{93},
032115 (2016).
\bibitem{Prabhu3} D. Sadhukhan, S. S. Roy, D. Rakshit, R. Prabhu, A. Sen(De)
and U. Sen, \textit{Quantum discord length is enhanced while entanglement length is not by introducing disorder in a spin chain},
Phys. Rev. E \textbf{93}, 012131 (2016).
\bibitem{Bera1} A. Bera, D. Rakshit, M. Lewenstein, A. Sen(De), U. Sen and
J. Wehr, \textit{Disorder-induced enhancement and critical scaling of
spontaneous magnetization in random-field quantum spin systems}, Phys. Rev. B \textbf{94}, 014421 (2016).
\bibitem{Bera2} A. Bera, D. Rakshit, A. Sen(De) and U. Sen, textit{Spontaneous
magnetization of quantum xy spin model in joint presence
of quenched and annealed disorder}, Phys. Rev. B \textbf{95}, 224441
(2017).
\bibitem{Aharony1} D. Stauffer and A. Aharony, \textit{Introduction to percolation theory}
(Taylor \& Francis, 2018).
\bibitem{Bera3} A. Bera, D. Sadhukhan, D. Rakshit, A. Sen(De) and U. Sen,
\textit{Response of entanglement to annealed vis-à-vis quenched disorder in quantum spin models}, EPL \textbf{127}, 30003 (2019).
\bibitem{Ghosh1} S. Ghosh, T. Chanda and A. Sen(De), \textit{Enhancement in the
performance of a quantum battery by ordered and disordered
interactions}, Phys. Rev. A \textbf{101}, 032115 (2020).
\bibitem{Ghoshal_Das} A. Ghoshal, S. Das, A. Sen(De) and U. Sen, \textit{Population inversion and entanglement in single and double glassy Jaynes Cummings models}, Phys. Rev. A \textbf{101}, 053805 (2020).
\bibitem{Sarkar_Sen} S. Sarkar and U. Sen, \textit{Glassy disorder-induced effects in noisy
dynamics of bose–hubbard and fermi–hubbard systems}, Journal of Physics B: Atomic, Molecular and Optical Physics \textbf{55}, 205502 (2022).
\bibitem{Tanoy} T. K. Konar, S. Ghosh, A. K. Pal and A. S.(De), \textit{Designing robust quantum refrigerators in disordered spin models},
Phys. Rev. A \textbf{105}, 022214 (2022).

\bibitem{Matsuzaki2} S. Kukita, Y. Matsuzaki and Y. Kondo, \textit{Heisenberg-limited quantum metrology using collective dephasing}, Phys. Rev. Appl. \textbf{16}, 064026 (2021).

\bibitem{An} K. Bai, Z. Peng, H.G. Luo and J.H. An, \textit{Retrieving Ideal Precision in Noisy Quantum Optical Metrology}
Phys. Rev. Lett. \textbf{123}, 040402 (2019).
\bibitem{Tong_Luo} Q.-J. Tong, J.-H. An, H.-G. Luo and C. H. Oh, \textit{Mechanism of entanglement preservation}, Phys. Rev. A \textbf{81},
052330 (2010).
\bibitem{Matsuzaki_new}Y. Matsuzaki, S. C. Benjamin and J. Fitzsimons, \textit{Magnetic field sensing beyond the standard quantum limit under the effect of decoherence}, Phys. Rev. A \textbf{84}, 012103 (2011).
\bibitem{Huelga3} A. W. Chin, S. F. Huelga and M. B. Plenio, \textit{Quantum
metrology in non-markovian environments}, Phys. Rev.
Lett. \textbf{109}, 233601 (2012).
\bibitem{Berrada} K. Berrada, \textit{Non-markovian effect on the precision of parameter estimation}, Phys. Rev. A \textbf{88}, 035806 (2013).
\bibitem{Xu} H.-B. Liu, W. L. Yang, J.-H. An and Z.-Y. Xu, \textit{Mechanism for quantum speedup in open quantum systems},
Phys. Rev. A \textbf{93}, 020105(R) (2016).
\bibitem{Lin} C.-J. Yang, J.-H. An and H.-Q. Lin, \textit{Signatures of quantized coupling between quantum emitters and localized
surface plasmons}, Phys. Rev. Res. \textbf{1}, 023027 (2019).
\bibitem{Shi} W. Wu and C. Shi, \textit{Quantum parameter estimation in
a dissipative environment}, Phys. Rev. A \textbf{102}, 032607
(2020).

\bibitem{clock1} J. E. Thomas, P. R. Hemmer, S. Ezekiel, C. C. Leiby, Jr., R. H. Picard, and C. R. Willis, \textit{Observation of Ramsey Fringes Using a Stimulated, Resonance Raman Transition in a Sodium Atomic Beam}
Phys. Rev. Lett. \textbf{48}, 867 (1982).
\bibitem{clock2} G. Santarelli, Ph. Laurent, P. Lemonde, A. Clairon, A. G. Mann, S. Chang, A. N. Luiten, and C. Salomon, \textit{Quantum Projection Noise in an Atomic Fountain: A High Stability Cesium Frequency Standard}
Phys. Rev. Lett. \textbf{82}, 4619 (1999).
\bibitem{clock3} T. Zanon, S. Guerandel, E. de Clercq, D. Holleville, N. Dimarcq, and A. Clairon, \textit{High Contrast Ramsey Fringes with Coherent-Population-Trapping Pulses in a Double Lambda Atomic System}
Phys. Rev. Lett. \textbf{94}, 193002 (2005).
\bibitem{clock4} Schioppo, M., Brown, R., McGrew, W. et al. \textit{Ultrastable optical clock with two cold-atom ensembles}, Nature Photon \textbf{11}, 48–52 (2017). 

%%%%%%%%%%%%%%%%%%%%%%%%%%%%%%%%%%%%%%%%%%%%%%%%%%%%%%%%%%%%%%%%%%%%%%%%%%%%%%%%
\bibitem{noise0}Z. Ji, G. Wang, R. Duan, Y. Feng and M. Ying,
 \textit{parameter estimation of quantum channels}, IEEE Trans. Inf. Theory, \textbf{54}, 11, 5172-5185, (2008).
 
\bibitem{Rafao_loss}U. Dorner, R. Demkowicz-Dobrza\'{n}ski, B. J. Smith, J. S. Lundeen, W. Wasilewski, K. Banaszek, and I. A. Walmsley, \textit{Optimal Quantum Phase Estimation}, Phys. Rev. Lett. \textbf{102}, 040403 (2009).

\bibitem{Rafao1}J. Kołodyński and R. Demkowicz-Dobrzański, \textit{Efficient tools for quantum metrology with uncorrelated noise}, New J. Phys. \textbf{15} 073043.

\bibitem{Rafao2}R. Demkowicz-Dobrzanski, U. Dorner, B. J. Smith, J. S. Lundeen, W. Wasilewski, K. Banaszek and I. A. Walmsley, \textit{Quantum phase estimation with lossy interferometers}, Phys. Rev. A \textbf{80}, 013825 (2009).

\bibitem{noise1}J. Kołody\'{n}ski and R. Demkowicz-Dobrza\'{n}ski, \textit{Phase estimation without a priori phase knowledge in the presence of loss}, Phys. Rev. A \textbf{82}, 053804 (2010).
\bibitem{noise2}S. Knysh, V. N. Smelyanskiy, and G. A. Durkin, \textit{Scaling laws for precision in quantum interferometry and the bifurcation landscape of the optimal state}, Phys. Rev. A \textbf{83}, 021804(R) (2011).
\bibitem{noise3}B. M. Escher, R. L. de Matos Filho and L. Davidovich, \textit{General framework for estimating the ultimate precision limit in noisy quantum-enhanced metrology}, Nat. Phys. \textbf{7}, 406–411 (2011).

\bibitem{decoher1}  B. M. Escher, R. L. de Matos Filho, and L. Davidovich, \textit{Quantum metrology for noisy systems}, 
Braz. J. Phys. \textbf{41}, 229 (2011).
\bibitem{decoher2}  B. M. Escher, L. Davidovich, N. Zagury, and R. L. de Matos Filho, \textit{Quantum Metrological Limits via a Variational Approach}, Phys. Rev. Lett. \textbf{109}, 190404 (2012).
\bibitem{decoher3} A. W. Chin, S. F. Huelga, and M. B. Plenio, \textit{Quantum Metrology in Non-Markovian Environments}, Phys. Rev. Lett.
\textbf{109}, 233601 (2012).

\bibitem{Rafao_dis1}R. Demkowicz-Dobrza\'{n}ski, J. Kołody\'{n}ski and M. Guta, \textit{The elusive Heisenberg limit in quantum-enhanced metrology}, 
 Nat. Com. \textbf{3}, 1063 (2012).
 
\bibitem{Rafao_dis2}R. Demkowicz-Dobrza\'{n}ski and J. Kołody\'{n}ski,  \textit{Efficient tools for quantum metrology with uncorrelated noise}, New J. Phys. \textbf{15}, 073043 (2013).

\bibitem{decoher4} R. Chaves, J. B. Brask, M. Markiewicz, J. Kolodynski, and A. Acin, \textit{Noisy Metrology beyond the Standard Quantum Limit}, Phys. Rev. Lett. \textbf{111}, 120401 (2013).

\bibitem{ref1} X.-X. Zhang, Y.-X. Yang, and X.-B. Wang, \textit{Lossy quantum-optical metrology with squeezed states}, 
Phys. Rev. A \textbf{88}, (2013).

\bibitem{decoher5} R. Demkowicz-Dobrza\'{n}ski and L. Maccone, \textit{Using Entanglement Against Noise in Quantum Metrology} Phys. Rev. Lett. \textbf{113}, 250801 (2014).
\bibitem{decoher6} W. Dür, M. Skotiniotis, F. Fröwis, and B. Kraus, Phys. Rev. Lett. \textit{Improved Quantum Metrology Using Quantum Error Correction}, \textbf{112}, 080801 (2014).

\bibitem{ref2}J. Jeske, J. H. Cole1 and S. F Huelga \textit{Quantum metrology subject to spatially correlated Markovian noise: restoring the Heisenberg limit} New J. Phys. \textbf{16} 073039, (2014).

\bibitem{Rafao3} R.Demkowicz-Dobrza\'{n}ski, M. Jarzyna, J.Ko\l{}ody\'{n}ski, \textit{Quantum Limits in Optical Interferometry}, Progress in Optics, \textbf{60}, (2015).

\bibitem{decoher7} A. Smirne, J. Kolodynski, S. F. Huelga, and R.
Demkowicz-Dobrzański, \textit{Ultimate Precision Limits for Noisy Frequency Estimation}, Phys. Rev. Lett. \textbf{116}, 120801
(2016).

\bibitem{Campo}M. Beau and A. del Campo, \textit{Nonlinear quantum metrology of many-body open systems}, Phys. Rev. Lett. \textbf{119}, 010403 (2017).
\bibitem{Campo2}M. Beau, J. Kiukas, I. L. Egusquiza and A. del Campo, \textit{Nonexponential Quantum Decay under Environmental Decoherence}, Phys. Rev. Lett. \textbf{119} 130401 (2017).

\bibitem{R_emni} R. Demkowicz-Dobrza\'{n}ski, J. Czajkowski and P. Sekatski, \textit{Adaptive Quantum Metrology under General Markovian Noise},
Phys. Rev. X \textbf{7} 041009 (2017).
\bibitem{emni1}F. Albarelli, M. A. C. Rossi, D. Tamascelli, and M. G. Genoni, \textit{Restoring Heisenberg scaling in noisy quantum metrology by monitoring the environment}, Quantum \textbf{2} 110 (2018).
\bibitem{emni2}D. P. Pires, I. A. Silva, E. R. deAzevedo, D. O. Soares-Pinto and J. G. Filgueiras, \textit{Coherence orders, decoherence, and quantum metrology},
Phys. Rev. A \textbf{98} 032101 (2018).
\bibitem{new0}F. Albarelli, M. A. C. Rossi and M. G. Genoni, \textit{Quantum frequency estimation with conditional states of continuously monitored independent dephasing channels}, Int. J. Quantum Inform. \textbf{18} 1941013 (2020).
\bibitem{new1}C. Chen, P. Wang and R.-B. Liu, \textit{Effects of local decoherence on quantum critical metrology}, Phys. Rev. A \textbf{104} L020601 (2021).

\bibitem{ref4}F. C.-Blondeau, \textit{Noisy quantum metrology with the assistance of indefinite causal order},
Phys. Rev. A \textbf{103}, 032615 (2021).

\bibitem{ref6}W. Wu and J.-H. An, \textit{Gaussian quantum metrology in a dissipative environment}, Phys. Rev. A \textbf{104}, 042609 (2021).

\bibitem{ref7}F. C.-Blondeau, \textit{Quantum parameter estimation on coherently superposed noisy channels}, Phys. Rev. A 104, 032214 (2021).

\bibitem{noise4}Y. L. Len, T. Gefen, A. Retzker and J. Kołody\'{n}ski, \textit{Quantum metrology with imperfect measurements}, Nat. Com. \textbf{13}, 6971 (2022).

\bibitem{ref5} Fr. Albarelli and R. Demkowicz-Dobrza\'{n}ski, \textit{Probe Incompatibility in Multiparameter Noisy Quantum Metrology}, 
Phys. Rev. X \textbf{12}, 011039 (2022).

%%%%%%%%%%%%%%%%%%%%%%%%%%%%%%%%%%%%%%%%%%%%%%%%%%%%%%%%%%%%%%%%%%%%%%%%%%%%%%%

%%%%%%%%%%%%%%%%%%%%%%%%%%%%%%%%%%%%%%%%%%%%%%%%%%%%%%%%%%%
%\bibitem{ref1} X.-X. Zhang, Y.-X. Yang, and X.-B. Wang, \textit{Lossy quantum-optical metrology with squeezed states}, 
%Phys. Rev. A \textbf{88}, (2013).

%\bibitem{ref2}J. Jeske, J. H. Cole1 and S. F Huelga \textit{Quantum metrology subject to spatially correlated Markovian noise: restoring the Heisenberg limit} New J. Phys. \textbf{16} 073039, (2014).

%\bibitem{ref3}Effects of local decoherence on quantum critical metrology
%Chong Chen, Ping Wang, and Ren-Bao Liu
%Phys. Rev. A 104, L020601 – Published 2 August 2021.

%%\bibitem{ref4}F. C.-Blondeau, \textit{Noisy quantum metrology with the assistance of indefinite causal order},
%Phys. Rev. A \textbf{103}, 032615 (2021).

%\bibitem{ref6}W. Wu and J.-H. An, \textit{Gaussian quantum metrology in a dissipative environment}, Phys. Rev. A \textbf{104}, 042609 (2021).

%\bibitem{ref7}F. C.-Blondeau, \textit{Quantum parameter estimation on coherently superposed noisy channels}, Phys. Rev. A 104, 032214 (2021).

%\bibitem{ref5} Fr. Albarelli and R. Demkowicz-Dobrza\'{n}ski, \textit{Probe Incompatibility in Multiparameter Noisy Quantum Metrology}, 
%Phys. Rev. X \textbf{12}, 011039 (2022).

%%%%%%%%%%%%%%%%%%%%%%%%%%%%%%%%%%%%%%%%%%%%%%%%%%%%%%%%%%%%

\bibitem{clock} S. Haroche, M. Brune and J.-M. Raimond, \textit{Atomic clocks for controlling light fields}, Physics Today \textbf{66}, 1, (2013).

%%%%%%%%%%%%%%%%%%%%%%%%
\bibitem{opt_lat1} V. Ahufinger, L. Sanchez-Palencia, A. Kantian, A. Sanpera, and M. Lewenstein, \textit{Disordered ultracold atomic gases in optical lattices: A case study of Fermi-Bose mixtures}, Phys. Rev. A \textbf{72}, 063616 (2005).

\bibitem{ultrac4} F. Crépin, G. Zaránd and P. Simon, 
\textit{Mixtures of ultracold atoms in one-dimensional disordered potentials}, Phys. Rev. A \textbf{85}, 023625 (2012).

\bibitem{ultrac1} M.  Lewenstein, A.  Sanpera, V.Ahufinger, B. Damski, A. Sen(De) and U. Sen, \textit{Ultracold atomic gases in optical lattices: mimicking condensed matter physics and beyond}, Adv. Phys., \textbf{56}, 243 (2007).

\bibitem{ultrac2} P. Lugan, D. Clement, P. Bouyer, A. Aspect, M. Lewenstein and L. Sanchez-Palencia, \textit{Ultracold Bose Gases in 1D Disorder: From Lifshits Glass to Bose-Einstein Condensate}, 
Phys. Rev. Lett. \textbf{98}, 170403 (2007).

\bibitem{ultrac3} F. Crépin, G. Zaránd and P. Simon, \textit{Disordered One-Dimensional Bose-Fermi Mixtures: The Bose-Fermi Glass}, Phys. Rev. Lett. \textbf{105}, 115301 (2010).
%%%%%%%%%%%%%%%%%%%%%%%%
\bibitem{matter_wave} Chen Li, Qi Liang, Pradyumna Paranjape, RuGway Wu, and Jörg Schmiedmayer, \textit{Matter-wave interferometers with trapped strongly interacting Feshbach molecules}, Phys. Rev. Research \textbf{6}, 023217 (2024).
%%%%%%%%%%%%%%%%%%%%%%%%
%\bibitem{Campo}M. Beau and A. del Campo, \textit{Nonlinear quantum metrology of many-body open systems}, Phys. Rev. Lett. \textbf{119}, 010403 (2017).


%\bibitem{Braunstein_0}S. L.Braunstein, C. M.Caves and G.J.Milburn, \textit{Generalized Uncertainty Relations: Theory, Examples, and Lorentz Invariance}, Ann. Phys. \textbf{247}, 135 (1996).

%\bibitem{Maccone} V. Giovannetti, S. Lloyd, L. Maccone, \textit{Advances in Quantum Metrology}, Nature Photonics \textbf{5}, 222 (2011).
%\bibitem{Rafao}R. Demkowicz-Dobrzanski, U. Dorner, B. J. Smith, J. S. Lundeen, W. Wasilewski, K. Banaszek, and I. A. Walmsley, \textit{Quantum phase estimation with lossy interferometers},
%Phys. Rev. A \textbf{80}, 013825 (2009).

%\bibitem{Braunstein} S. L. Braunstein and C. M. Caves, \textit{Statistical distance and the geometry of quantum states
%}, Phys. Rev. Lett. \textbf{72}, 3439
%(1994).
%\bibitem{Rafao1}J. Kołodyński and R. Demkowicz-Dobrzański, \textit{Efficient tools for quantum metrology with uncorrelated noise},
% New J. Phys. \textbf{15} 073043.
% \bibitem{Rafao2}R. Demkowicz-Dobrzanski, U. Dorner, B. J. Smith, J. S. Lundeen, W. Wasilewski, K. Banaszek and I. A. Walmsley, \textit{Quantum phase estimation with lossy interferometers},
%Phys. Rev. A \textbf{80}, 013825 (2009).
%\bibitem{Rafao3} R.Demkowicz-Dobrza\'{n}ski, M. Jarzyna, J.Ko\l{}ody\'{n}ski, \textit{Quantum Limits in Optical Interferometry}, Progress in Optics, \textbf{60}, 2015.

%\bibitem{ramsey} C. Gardiner and P. Zoller, \textit{The quantum world of ultra-cold atoms and light book II: The physics of quantum-optical devices}, (Imperial College Press, 2015).

\bibitem{Chowdhury}D. Chowdhury, \textit{Spin Glasses and Other Frustrated Systems},
(World Scientific Publishing, Singapore, 1986).

\bibitem{glassy1}A. Prem, J. Haah and R. Nandkishore, \textit{Glassy quantum dynamics in translation invariant fracton models}, Phys. Rev. B \textbf{95}, 155133 (2017).

\bibitem{glassy2}A. Signoles, T. Franz, R. Ferracini Alves, M. Gärttner, S. Whitlock, G. Zürn and M. Weidemüller, \textit{Glassy Dynamics in a Disordered Heisenberg Quantum Spin System}, Phys. Rev. X \textbf{11}, 011011 (2021).

\bibitem{glassy3}A. Braemer, T. Franz, M. Weidemüller and M. Gärttner, \textit{Pair localization in dipolar systems with tunable positional disorder}, Phys. Rev. B \textbf{106}, 134212 (2022).

\bibitem{Parisi} M. Mezard, G.Parisi and M. Virasoro, \textit{Spin Glass Theory and Beyond: An Introduction to the Replica Method and Its Applications}, (World Scientific Publishing, Singapore, 1987). 
\bibitem{Dutta} B. K. Chakrabarti, A. Dutta and P. Sen,\textit{ Quantum Ising Phases and Transitions in Transverse Ising Models}, (Springer, Berlin, 1996).
\bibitem{Nishimori} H. Nishimori, \textit{Statistical Physics of Spin Glasses and Information Processing An Introduction}, (Clarendon Press, Oxford, 2001).
\bibitem{Giardina} C. De Dominicis and I. Giardina, \textit{Random Fields and Spin
Glasses: A Field Theory Approach} (Cambridge University
Press, 2006).
\bibitem{Sachdev}S. Sachdev, \textit{Quantum Phase Transitions}, (Cambridge University Press, Cambridge, 2011);
\bibitem{Suzuki} S. Suzuki, J.-I. Inoue, B. K. Chakrabarti, \textit{Quantum Ising Phases and Transitions in Transverse Ising Models}, (Springer, Berlin, 2013).


\bibitem{median1} W. Feller, \textit{An Introduction to Probability Theory and its Applications}, (Wiley, New York, 1968). 
\bibitem{median2} A. Gupta, \textit{Groundwork of Mathematical Probability and Statistics}, (Academic Publishers, Kolkata, 2012).

%\bibitem{root} To solve $x_{M_G}$,  we have used $\tan^{-1}a+\tan^{-1}b=\tan^{-1}(a+b)/(1-ab)$, for $-1 < ab \le 1$.

\bibitem{footnote} To solve 
$x_{M_{CL}}$,  we have used 
$\tan^{-1}a+\tan^{-1}b=\tan^{-1}(a+b)/(1-ab)$, for $-1 < ab \le 1$.

\bibitem{Bhattacharyya} A. Bhattacharyya, A. Ghoshal and U. Sen, \textit{Restoring metrological quantum advantage of measurement precision in noisy scenario},  	arXiv:2211.05537.


\bibitem{Bennett_Popescu}C. H. Bennett, H. J. Bernstein, S. Popescu and Benjamin Schumacher, \textit{Concentrating partial entanglement by local operations},
Phys. Rev. A \textbf{53}, 2046 (1996).

\bibitem{Sanderson}  C. Sanderson and R. Curtin, \textit{Armadillo: a template-based C++ library for linear
algebra}, Journal of Open Source Software 1, 26 (2016).
\bibitem{Sanderson1} C. Sanderson and R. Curtin, Lecture Notes in Computer Science (LNCS) 10931, 422 (2018).
\bibitem{QIClib} T. Chanda, \emph{QIClib}, \url{https://titaschanda.github.io/QIClib}.

\end{thebibliography}
\end{document}